\documentclass[twocolumn]{aastex631}

\shorttitle{Short GRB Afterglows}
\shortauthors{Castrejon et al.}

\begin{document}

\title{Exploring the Relationship Between {\it Swift} Short Gamma-Ray Burst Afterglows and their Host Galaxy Properties}

\author[0009-0005-4757-8285]{Cristian Castrejon}
\affiliation{Center for Interdisciplinary Exploration and Research in Astrophysics (CIERA) and Department of Physics and Astronomy, Northwestern University, Evanston, IL
60208, USA}

\author[0000-0002-2028-9329]{Anya~E.~Nugent}
\affiliation{Center for Astrophysics | Harvard \& Smithsonian, 60 Garden St. Cambridge, MA 02138, USA}

\author[0000-0002-7374-935X]{Wen-fai Fong}
\affiliation{Center for Interdisciplinary Exploration and Research in Astrophysics (CIERA) and Department of Physics and Astronomy, Northwestern University, Evanston, IL
60208, USA}

\author[0000-0001-9915-8147]{Genevieve~Schroeder}
\affiliation{Department of Astronomy, Cornell University, Ithaca, NY 14853, USA}

\author[0000-0003-3937-0618]{Alicia~Rouco~Escorial}
\affiliation{European Space Agency (ESA), European Space Astronomy Centre (ESAC), Camino Bajo del Castillo s/n, 28692 Villanueva de la Cañada, Madrid, Spain}

\author{Olivia~Guerra}
\affiliation{Center for Interdisciplinary Exploration and Research in Astrophysics (CIERA) and Department of Physics and Astronomy, Northwestern University, Evanston, IL
60208, USA}

\begin{abstract}

We present a comprehensive compilation of short-duration gamma-ray burst (GRB) afterglows in the X-ray, optical, and radio bands, comprising 150 events discovered primarily by the Neil Gehrels {\it Swift} Observatory over 2005-2023. We pair these observations with uniformly modeled host galaxies to understand how broadband afterglow luminosities are influenced by their environmental properties. We compare the X-ray and optical afterglow luminosities at $3$~hr with projected physical and host-normalized galactocentric offsets, host stellar mass, star-formation rate (SFR), specific SFR, and stellar population age. In the radio band, we explore how these environmental properties may influence afterglow detectability. We find statistical support that X-ray afterglows are brighter in galaxies with younger ages, lower masses, and higher active star formation - trends that also scale with ISM density. While we also visualize these differences for optical afterglows, the only statistically significant trend is that they are brighter in hosts with higher SFR. We further find that X-ray (radio) afterglows are more luminous (more likely to be detected) at low projected offsets. Overall, this indicates that X-ray afterglow luminosity is the most predictable indicator of host environment among the three bands. We find the afterglow luminosities of three possible merger-driven long GRBs to be unremarkable compared to the traditional short GRB population, strengthening the case that these events arise from mergers. Finally we find that the estimated on-axis afterglow luminosity of GW170817 is in the faintest $\approx 30\%$, aligning with its quiescent, old and massive host environment. 
\end{abstract}

\keywords{short gamma-ray bursts, galaxies}

\section{Introduction} \label{sec:intro}
The relationship between properties of astrophysical transients and their host galaxies has long been leveraged to disentangle observed features that may be due to the progenitor systems from effects due to the environment. For example, Type Ia supernovae (SNe~Ia), which derive from degenerate progenitors and are vital probes of the expansion rate of the Universe \citep{riess1998,perlmutter1999}, are observed to have brighter peak luminosities in host galaxies with higher stellar masses ($\gtrsim10^{10} M_\odot$), lower specific star formation rates (sSFR), and higher gas-phase metallicities than those in the low mass, high sSFR, and low metallity hosts \citep{sullivan2006,neill2009,kelly2010,  lampeitl2010, sullivan2010,dgs+2011,gupta2011, childress2013, pan2014, rigault2020}. It is thought that these observed trends arise from differences in dust relations between high- and low-mass galaxies; thus, the SN luminosity versus host property trends are a result of the environment as opposed to a difference in progenitor systems \citep{salim2018, bs2021, meldorf2023}. Trends with transient brightness and host properties have also been observed for the long-duration $\gamma$-ray burst (GRB) population, which are highly energetic, jetted explosions of $\gamma$-ray emission that last for $\gtrsim 2$~seconds \citep{Norris1984, kmf+1993, Dezalay1992, Lien+2016} and derive from massive stars. Long GRBs are followed by multi-wavelength synchrotron `afterglow' emission (e.g., \citealt{RM1992}) that results from the interaction of the relativistic jet with the circumburst environment. A subset of long GRBs with detected X-ray (and often radio) afterglows, have fainter optical afterglows than expected and often evade detection. Although such long GRBs, coined ``dark" \citep{Fynbo2001}, are not distinct in their intrinsic $\gamma$-ray properties from optically-bright GRBs \citep{Schroeder2022}, they tend to occur in hosts with higher global dust extinction and stellar masses than those with detected optical afterglows \citep{Kruhler2011,Perley2009,Perley2013,Schroeder2022}. Just as in SNe Ia, this demonstrates that the fainter optical afterglows of dark GRBs are largely a product of their environments, as opposed to an intrinsic trait of the transient or its progenitor. 

A related class of high-energy transients are short-duration GRBs (nominal durations $\lesssim 2$~sec), the majority of which likely originate from the mergers of two neutron stars, or a neutron star and a black hole \citep{Abbott+2017-grb,berger2014}. The afterglows of short GRBs are $\approx 10-100$ times fainter across the electromagnetic spectrum than those of long GRBs \citep{Panaitescu+2001,Berger+2007-afterglows,Nakar2007,Kann2011}. One natural explanation is that short GRBs have intrinsically lower kinetic energy scales (e.g., \citealt{Panaitescu+2001}), since the brightness of the afterglow scales with kinetic energy at all wavelengths \citep{Granot+2002}. Accurate inference on the kinetic energy relies on well-sampled afterglows to enable robust modeling, as well as constraints on collimation to correct from isotropic-equivalent values to the true ones \citep{Rhoads1999}. Indeed, based on such studies for a $\sim$dozen short GRBs, the intrinsic, beaming-corrected kinetic energies of short GRBs are $\approx 10^{49}-10^{50}$~erg \citep{Fong+2015,rfb+2023}, $\sim 1-2$ orders of magnitude lower than for long GRBs (e.g., \citealt{gehrels+2008}). However, some short GRBs have evidence for wider jets \citep{Laskar+2022,Schroeder+2024,Schroeder+2025,RoucoEscorial2023}, signifying larger energy scales that may not be able to explain the faintness of short GRB afterglows alone. 

An additional possibility is that short GRBs arise in environments with lower circumburst densities, a parameter which also scales with afterglow brightness. Indeed, while long GRBs are generally located proximate to star formation regions within their star-forming host galaxies \citep{Perley2013, Wang2014, Vergani2015, Blanchard+2016,Niino2017}, short GRBs occur in both actively star-forming and quiescent galaxies \citep{LeiblerBerger+2010, Nugent+2022,Jeong+2022} and also occur in more remote locations \citep{cld+11, fb13, tlt+14, Fong+2022, O'Conner+2022}. These characteristics of the short GRB population are naturally explained by their neutron star (NS) binary progenitors, which experience a combination of systemic velocities as observed with the Galactic population \citep{tkf+17, vns+18, am19, gaspari2024} and substantial delay times of $\sim 10$'s of Myr to few Gyr \citep{Nakar2007, Berger+2007, Jeong2010, Hao2013, Wanderman2015, tkf+17, Anand2018, am19, Zevin+DTD}. As a result, short GRBs arise in older stellar populations and exhibit locations more offset from star formation than long GRBs. Commensurately, short GRBs have lower circumburst densities ($10^{-3}-10^{-2}$~cm$^{-3}$; \citealt{Soderberg+2006,Fong+2015,O'Connor+2020,RoucoEscorial2023}) than long GRBs ($\approx0.1-100$~cm$^{-3}$; \citealt{Schneider+2022, Yost+2003,Chrimes+2022}). A recent study showed that the small but growing population of short GRBs with radio afterglow detections have larger inferred densities than those without detections \citep{Schroeder+2025}. Taken together, density very likely plays a role in the observed afterglow luminosity differences between long and short GRBs (e.g., \citealt{PernaBelcynzski2002}).

However, the trends between afterglow brightness with kinetic energy and densities are based on subsets of the population with relevant data, and inferred values for these parameters are subject to modeling differences. This has motivated studies which rely on observed, as opposed to inferred, properties for larger numbers of short GRBs. For instance, \citet{O'Conner+2022} investigated the ratio of X-ray flux to $\gamma$-ray fluence (a proxy for density if the cooling frequency is above the X-ray band; \citealt{Granot+2002}) versus projected galactocentric offset for a few dozen events and did not find any clear trends. In the radio band, short GRBs with radio afterglow detections tend to have smaller offsets than those which evade detection, although statistical significance in this trend has not been established \citep{Schroeder+2025}. Finally, \citealt{Nugent+2022} found that short GRB optical afterglows may be brighter in galaxies with higher active star formation (which, at a basic level, correlates with ISM density). However, the visible trend was also not statistically significant.

A natural extension of these studies are to investigate the relationship between afterglow brightness and any host galaxy property (e.g., star formation rate, stellar mass, stellar population age, dust extinction), which will give deeper insight into how much environment plays a role in afterglow luminosity. However, this is challenging to conduct in a uniform way. First, the relatively low afterglow detection rates, particularly in the optical and radio bands, mean that a large number of upper limits need to be accounted for in any statistical study. Second, the diversity of sampling in light curves (with some having only one data point, and some having many), coupled with the rapid fading behavior, makes interpolation to a common time highly nuanced. Finally, until recently, a homogeneous set of short GRB host galaxy properties, including those events which lacked precise (sub-arcsecond) localization, did not exist.

\begin{figure*}[t]
\centering
\includegraphics[width=\textwidth]{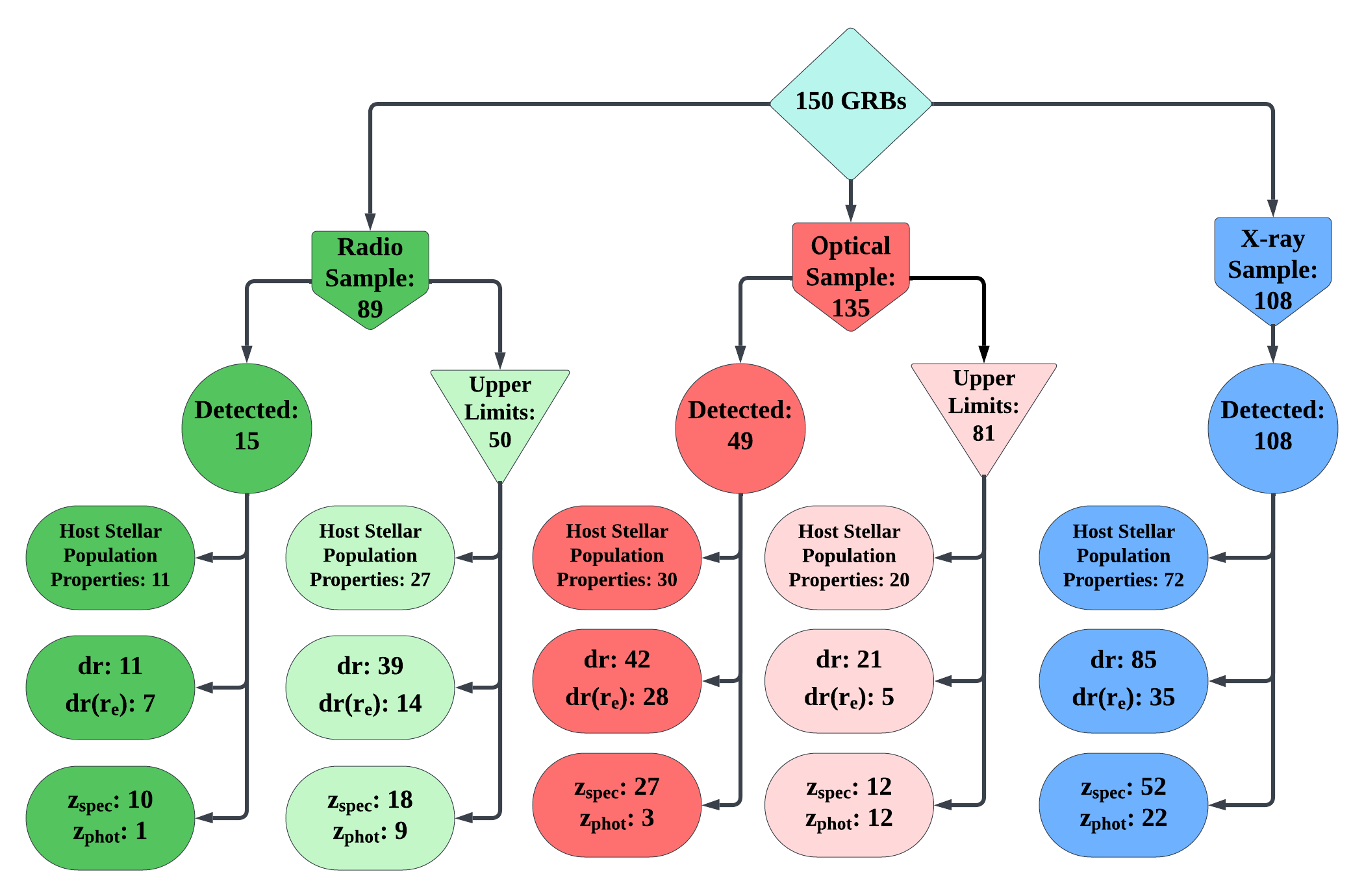}
\caption{A flowchart depicting the breakdown of our sample at each wavelength (X-ray: blue, optical: red, and radio: green). Within each wavelength, we display the number of GRBs that have either detected afterglows or upper limits. We further include the number of GRBs that have host stellar populations, physical and host-normalized offsets, and spectroscopic versus photometric redshifts. We note that five GRBs have no $griz$ or $V$-band filters available, and thus appear in the second row but not the third. Similarly, 24 GRBs with only shallow radio limits ($>100\,\mu$Jy) and otherwise no detections and are not considered in our radio upper limits.}
\label{fig:sample}
\end{figure*}

Here, we attempt to overcome these challenges and leverage the first substantial and uniformly-modeled samples of host galaxies, combined with two decades of short GRB afterglow observations across the electromagnetic spectrum. We analyze a population of 150 GRBs detected with the Neil Gehrels \textit{Swift} Observatory (\textit{Swift}) to investigate possible relations between X-ray, optical and radio afterglows to their host galaxy properties. In Section~\ref{sec:sample}, we describe our sample and identify the number of events with reported X-ray, optical, and radio afterglow data, host galaxy associations, redshifts, and host stellar population properties. In Section~\ref{sec:methods}, we detail our method to determine afterglow luminosity at a specified common rest-frame time. In Section~\ref{sec:results}, we investigate trends in the luminosities of X-ray and optical afterglows, as well as the detectability of radio afterglows, to environmental properties. We discuss implications for these findings and relate them to afterglow properties of long GRBs and gravitational wave events with NS merger origins in Section \ref{sec:discussion}. Finally, we list our main conclusions in Section \ref{sec:conclusion}.

Unless otherwise stated, all observations are reported in the AB magnitude system and have been corrected for Galactic extinction in the direction of the GRB \citep{MilkyWay,sf11}. We employ a standard WMAP9 cosmology of $H_{0}$ = 69.6~km~s$^{-1}$~Mpc$^{-1}$, $\Omega_{\rm m}$ = 0.286, $\Omega_{\rm vac}$ = 0.714 \citep{Hinshaw2013, blw+14}.

\section{GRB Sample} \label{sec:sample}

\begin{figure*}
\centering
\includegraphics[width=0.49\textwidth]{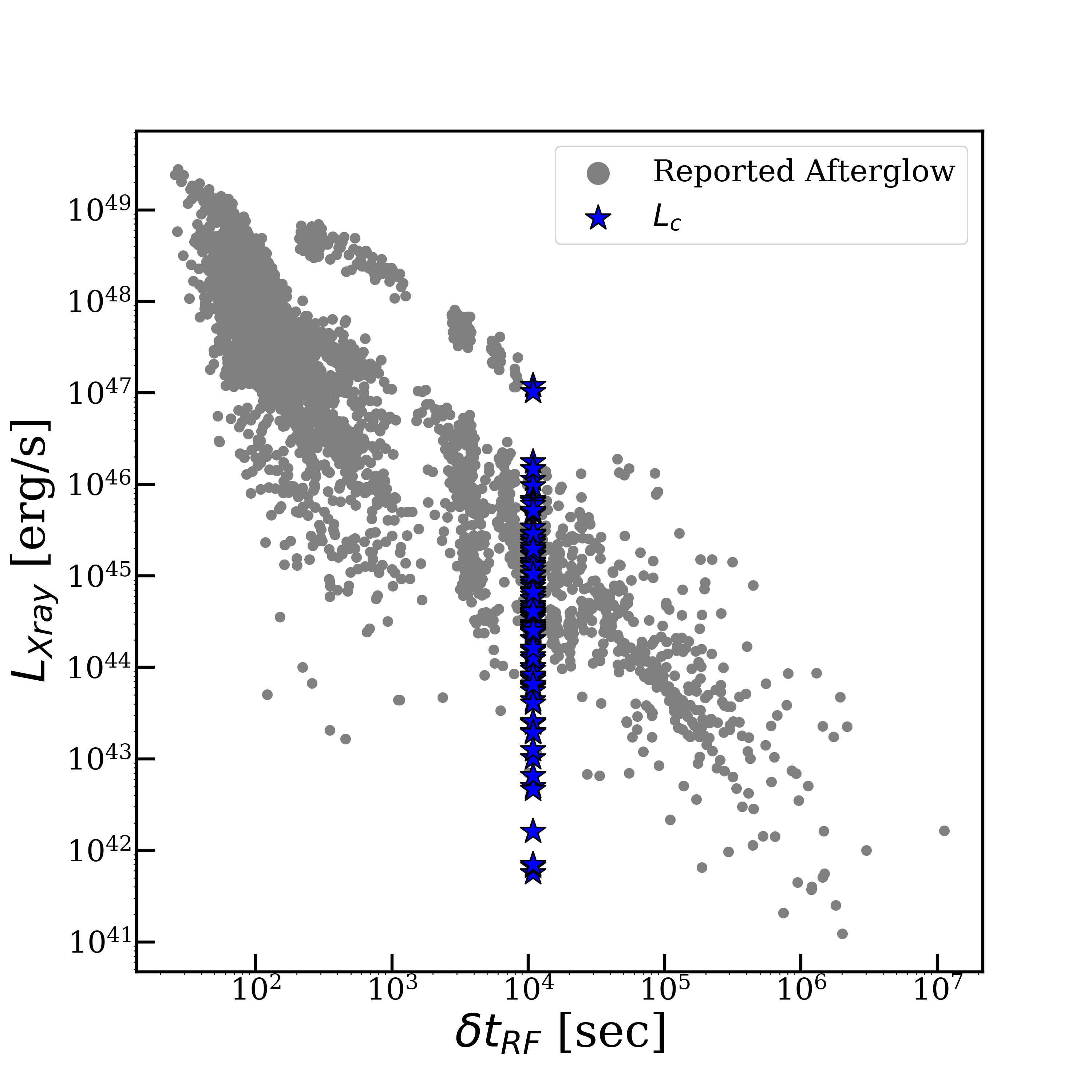}
\includegraphics[width=0.49\textwidth]{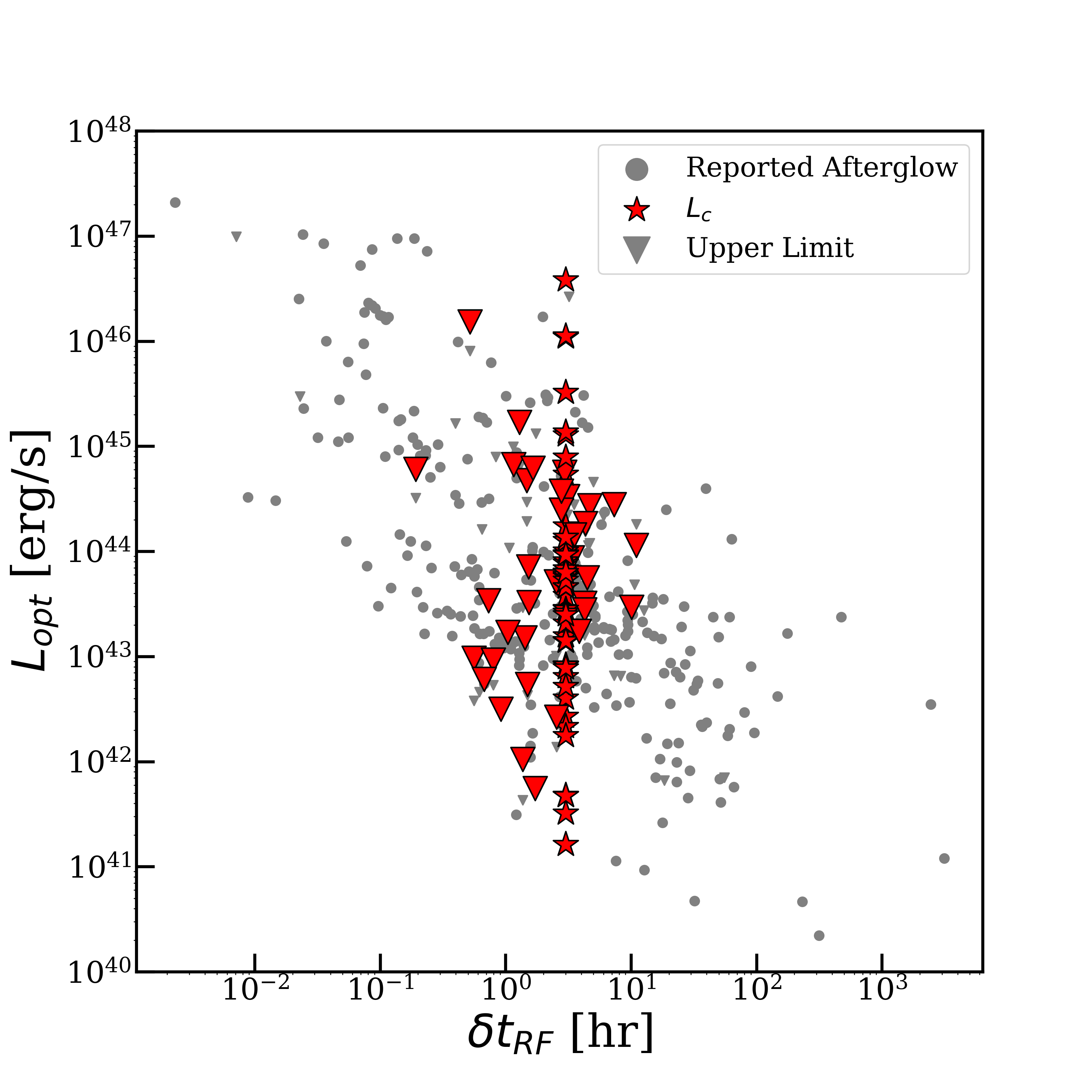}
\caption{\textit{Left:} The detected X-ray afterglow luminosities compared to rest-frame time after the burst ($\delta t_{\rm RF}$ for our short GRB sample (gray circles). The blue stars represent the luminosity determined at a common rest-frame time ($L_c$) of $3$~hr, following the methods described in Section \ref{sec:methods}. \textit{Right}: The detected optical afterglows (gray circles) and upper limits (gray triangles) for our short GRB sample. The red stars show the $L_c$ determined at $3$~hr and the red triangles show the GRBs with reported upper limits at $\delta t_{\rm RF} \pm 2.8$~hr which we use in our study.}
\label{fig:lum_sample}

\end{figure*}

We begin with the sample of all short GRBs detected by {\it Swift} \citep{ggg+04} with $T_{90}<$~2 seconds (excluding GRB 170817A), and those that have been classified as having extended emission \citep{Lien+2016}. We exclude any short GRBs that derive from collapsar origins (GRBs 080913, 090423, 100724A, and 200826A; \citealt{zhang2009, ahumada2021, rhodes2021, zhang2021}), as these likely do not share the same origin as the rest of the short GRB population. This comprises 150 GRBs that were observed and localized between February 2005 and November 2023 (ending with GRB\,231117A). Since our study hinges upon correlations with afterglow brightness, we trim the sample to only include those with any afterglow follow-up observations. This comprises 105 short GRB events in the X-ray band, 132 in the optical band, and 86 in the radio band. We also include three long GRBs: GRBs\,060614 \citep{Jin+2015, Yang+2015, Gompertz+2018, Rossi+2020}, 211211A \citep{Rastinejad+2022, Troja+2022, Yang+2022}, and 230307A \citep{Gillanders+2023, Levan+2024, Yang+2024}. Despite their long durations, these events are associated with probable $r$-process kilonovae (KN), indicative of NS merger origins. All have reported X-ray, optical, and radio afterglows. Thus, in total we consider 108 events with X-ray, 135 with optical, and 89 with radio observations. The summary of our sample, split by availability of observations within each wavelength, is depicted in Figure~\ref{fig:sample} and listed in Table~\ref{tab:wavelength_availibility}.

We collect all X-ray afterglow data from the \textit{Swift} light-curve repository \citep{evans2007,evans2009}, supplemented by \textit{Chandra} and XMM-Newton data reported in \citet{rfb+2023} and references therein. We collect optical afterglow data for 111 GRBs from \citet{Fong+2015}, \citet{Rastinejad+2021}, and references therein. For an additional 24 GRBs, we collect optical afterglow data from the Gamma-ray Coordination Network (GCN) stream and the literature. Of the bursts with optical data, we exclude five events that do not have any data in one of the $grizV$-bands. We find available radio afterglow data for 89 short GRBs within various GCNs and publications (see Table~\ref{tab:wavelength_availibility} for references). Of the 89 in the radio band, 24 have shallow radio observations of $\gtrsim 100\,\mu$Jy \citep{Chastain+2024, Schroeder+2025} resulting in non-detections; we do not include these GRBs in our sample of upper limits since we do not have meaningful constraints on their non-detection (see Table \ref{tab:wavelength_availibility}). The full list of references for our afterglow data is in Table~\ref{tab:wavelength_availibility}.

In addition to the afterglow data, we also require a host galaxy association for inclusion in the sample as our study is focused on correlations to host properties. All host associations, galactocentric offsets, redshifts, and host stellar population properties were obtained from the Broadband Repository for Investigating Gamma-ray burst Host Traits (BRIGHT\footnote{\url{http://bright.ciera.northwestern.edu}}; \citealt{Fong+2022, Nugent+2022}) and from \citet{Levan+2024, Schroeder+2025, Nugent+2025}.

In Figure \ref{fig:sample}, we display the total sample, categorized by their detectability in each wavelength. We further list the sample sizes for the subsets of bursts that have specific environmental properties. We show the X-ray and optical afterglow light curves of all events with detections and upper limits in Figure~\ref{fig:lum_sample} (see Section \ref{sec:methods}). In Table~\ref{tab:wavelength_availibility}, we list the $\gamma$-ray T$_{90}$ for each GRB and whether it has available X-ray, optical, and radio afterglow data (upper limits or detections). In total, we have collected 3812 X-ray detections, 1412 optical detections and upper limits, and 253 radio detections and upper limits. Our short GRB sample further spans a wide range of redshifts: $0.01 < z < 2.79$. This dataset is the largest collection of short GRB afterglows to-date.

\section{Methods} \label{sec:methods}

We seek to investigate whether short GRB afterglow luminosities are correlated with any environmental property, including the short GRB galactocentric offsets and their host galaxy stellar population properties. Given that our sample spans a wide range of redshifts (Section \ref{sec:sample}) and that afterglows fade rapidly, it is important to determine the afterglow luminosities ($L_c$) at a single common rest-frame time ($\delta t_c$) \citep{D'Avanzo+2012}. In the following subsections, we describe our methods for computing $L_c$ from the observed X-ray and optical afterglows. Owed to the very sparse sampling of radio afterglows, and the fact that radio afterglows rise at later times \citep{Granot+2002}, precluding simple interpolations, we do not calculate $L_c$ for radio afterglows. We do, however, present results based on whether or not a radio afterglow is detected (Section \ref{sec:results}).

\subsection{Deriving X-ray Luminosities at a Common Time}
\label{sec:xcommon}

In the X-rays, we first convert the afterglow fluxes to luminosities given the redshifts of each short GRB\footnote{Since the majority of our sample has X-ray detections, we do not use X-ray afterglow upper limits in this analysis.}. We then convert all observed afterglow times to the rest-frame to luminosities given the redshifts of each short GRB (Table \ref{tab:L_crf_results}). In cases in which the redshift is not known, we use the median redshift of the sample determined of $z\approx 0.64$ \citep{Fong+2022,Nugent+2022} for these conversions. We consider two different scenarios to determine $L_c$ depending on the amount of available data we have: (i) there exists $>1$~detection at $\delta t_c \pm 2.83$~hr\footnote{We find that this time interval is large enough so it includes enough data for accurate interpolation (which requires at least two data points) but small enough that it mitigates the effects of temporal breaks (e.g., \citealt{Rhoads1999}).}, or (ii) there are $\leq 1$~detections at $\delta t_c \pm 2.83$~hr. Hereafter, we adopt a value of $\delta t_c = 3$~hr, given that the majority of GRBs (71\%) have X-ray afterglow data observed within 2.83 hr of the $\delta t_c$. We also perform the same analysis at $\delta t_c=10$~hr which significantly trims the sample; in Section \ref{sec:results}, we comment on how our results are generally not sensitive to our choice.

To determine $L_c$ for scenario (i), we use \texttt{interp1d} from the Python package \texttt{scipy} to interpolate over the available luminosity values that fall within $\delta t_c \pm 2.83$~hr. We require that there exist at least one detection each, prior to and after $\delta t_c$, for proper interpolation. Furthermore, we do not include any X-ray data immediately following the burst ($\delta t \lesssim $~10~sec) as this often considered a feature from the burst itself rather than real afterglow \citep{nousek2006, D'Avanzo+2012}. We find that 55 GRBs fit the criteria for scenario (i). To determine uncertainties on $L_c$, we take all available X-ray data and find that the median uncertainty is $\sigma_{L_X}  /L_{X} \approx 0.2$; thus, we adopt 20\% of each $L_c$ as the fiducial uncertainty. 

For scenario (ii) ($\leq 1$ detections at $\delta t_c\pm2.83$~hr), we find that 53 GRBs fall in this category. For the GRBs with only one detection within 2.83~hr of $\delta t_c$, we take that detection and extrapolate to $\delta t_c=3$~hr, using a single power-law with $L_X \propto t^{-1}$ to calculate $L_c$. We employ a power-law index of $-1$ as this characterizes the median afterglow decline rate for short GRBs with well-sampled afterglows \citep{Margutti+2013, Fong+2015, RoucoEscorial2023}. As with scenario (i), we adopt $\sigma_{L_c} = 0.2  L_c$ as the uncertainty on $L_c$. For events with no detections within $\delta t_c\pm2.83$~hr, we use the nearest (in time) data point outside of this time interval, and perform the same extrapolation using $L_X \propto t^{-1}$. Of the 53 GRBs that fall in scenario (ii), only 14 are in this sub-category. We highlight the observed X-ray lightcurves and $L_X$ determined at $t_c = 3$~hours in Figure \ref{fig:lum_sample}.

\subsection{Deriving Optical Luminosities at a Common Time}

There a few key differences between optical and X-ray afterglow observations for short GRBs that require modification of our techniques for determining $L_c$. For example, KNe are an additional source of emission that can affect the behavior of optical afterglows and sometimes dominate the light at a given time or wavelength. For the 12 GRBs that have possible KNe\footnote{These are GRBs\,050709, 050724A, 060614, 070714B, 070809, 080503, 130603B, 150101B, 160821B, 200522A, 211211A, and 230307A.} we determine which detections are strongly contaminated by KN light based on published data \citep{Berger+2013, Tanvir+2013, Jin+2015, Yang+2015, Fong+2016, Jin+2016, Kasliwal+2017, Gompertz+2018, Troja+2018, Rossi+2020, Fong+2021, Rastinejad+2021, O'Conner+2022, Troja+2022, Yang+2022, Gillanders+2023, Zhou+2023, Levan+2024, Yang+2024}, and exclude all such data from our calculation of $L_c$. For GRBs\,050709 and 070809, if they truly have KNe, then all of the available optical data would be KN-dominated. However, the luminosities of the early-time points are also consistent with afterglow emission and their identification as KNe are less certain; thus, we only use the earliest observations for these two events. For GRB\,211211A, there is a well-sampled light curve with the contributions of the afterglow and kilonova determined self-consistently in \citet{Rastinejad+2025}. Thus, we use the afterglow contribution at $\delta t_c$ for this event. 
For GRB\,230307A, we use optical afterglow observations at $\delta t<24$~hr only, as observations beyond this clearly deviate from the afterglow models \citep{Rastinejad+2025,Yang+2024}. We mention the KN-dominated times ranges for these 12 events at the end of Table~\ref{tab:wavelength_availibility}.

As optical afterglow luminosity not only changes as a function of time, but also wavelength, we prioritize minimizing the number of filters used to derive $L_c$. Since the majority of our sample and environmental properties are $r$-band observations (52\%; including $r$', $R$, $R_c$, and $F606W$), we select this as the default filter. If no $r$-band data is available, we then prioritize the filters in the order of: $i$ (incl. $i', I, I', Ic$; 21\%), $g$ (incl. $g'$; 11\%), $z$ (incl. $z'$; 4\%), and $V$ (12\%). We correct all optical afterglow magnitudes for Galactic extinction in the direction of each burst and host galaxy extinction ($A_V$) if known; the results are in Table \ref{tab:L_crf_results}. 

After making a filter selection, we convert the optical afterglow light curves to rest-frame time and luminosities, using the same redshift considerations as we did for the X-rays (Section~\ref{sec:xcommon}). To compute $L_c$, we again employ different methods based on the number of single-filter optical afterglow detections that exist: there are (i) $\ge2$ detections, (ii) 1 detection, or (iii) no detections (only upper limits). We choose $\delta t_c=3$~hr, given that most upper limits, which cannot be interpolated, are near this time. Further, $\sim55\%$ of GRBs with $r$-band data fall within $\delta t_c\pm2.5$~hr. To address any potential bias in our conclusions by using $\delta t_c=3$~hr, we also perform an analysis at a later time of $\delta t_c=10$ hr, using the same methods to determine $L_c$ as described here. The majority of our conclusions remain when using either of these $\delta t_c$ and we point the reader to Section \ref{sec:results} for more details. 

We determine the $L_c$ for scenario (i), when there are $\geq2$ detections, by fitting the luminosities and rest-frame times with a power law in linear space, with the form: $L = n \times t^{\alpha}$. We employ the \texttt{curve\_fit} function from \texttt{scipy} to determine the optimal values for $n$ and $\alpha$, which uses a non-linear least-squares to fit the detected luminosity points to the power law. We then calculate $L_c$ using the optimal $n$ and $\alpha$. There are 34 GRBs that fit the criteria for this scenario. In the case when there are more than two data points, we calculate $\sigma_{L_C}$ using the estimated approximate covariance matrix output from the \texttt{curve\_fit} function. This applies to 21 out of the 34 GRBs. In cases where there are only two data points, preventing an accurate error on $L_c$, we instead use $\sigma_{L_c}$ = $0.15  L_c$ (the median optical afterglow luminosity error). 

\begin{figure*}[t]
\centering
\includegraphics[width=0.3\textwidth]{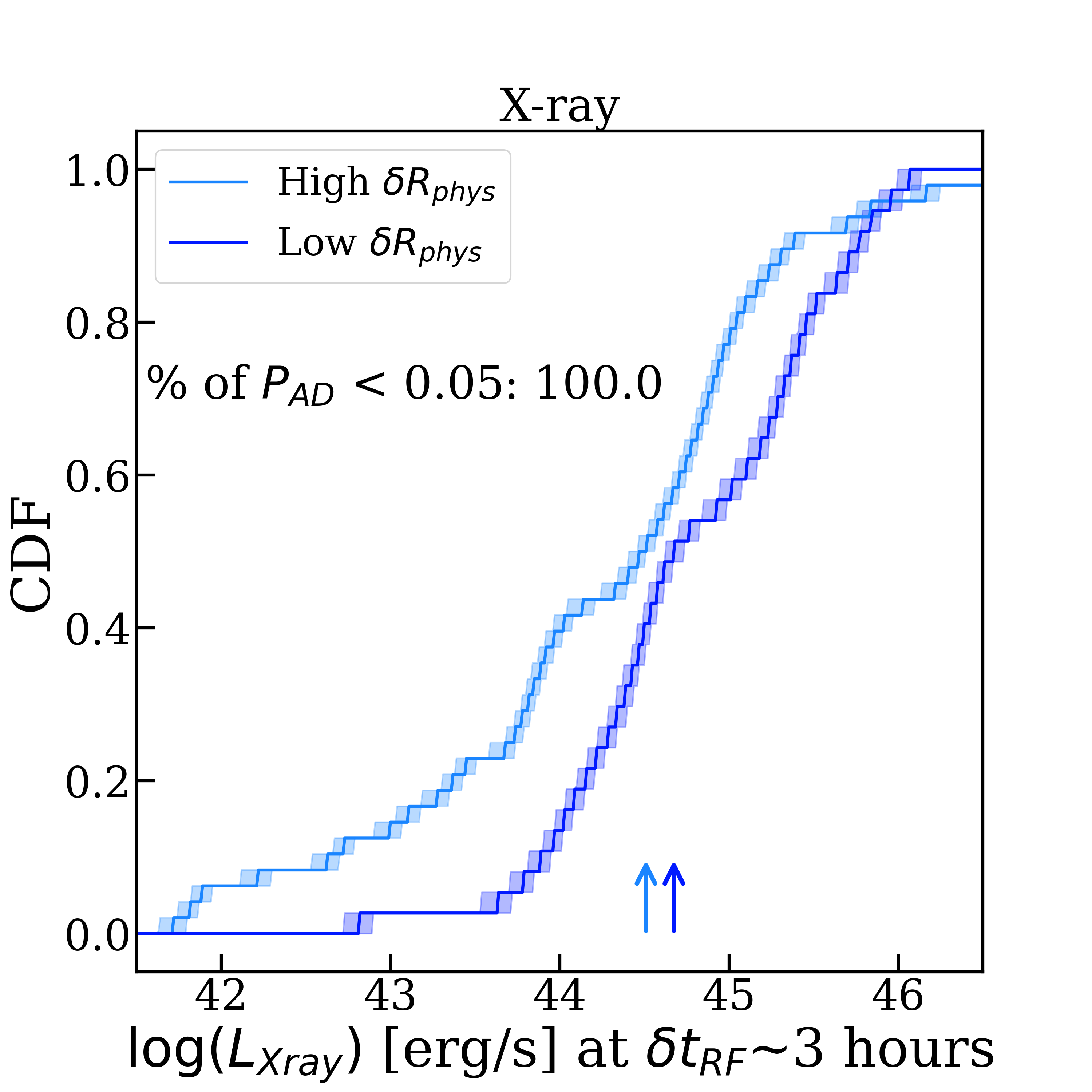}
\includegraphics[width=0.3\textwidth]{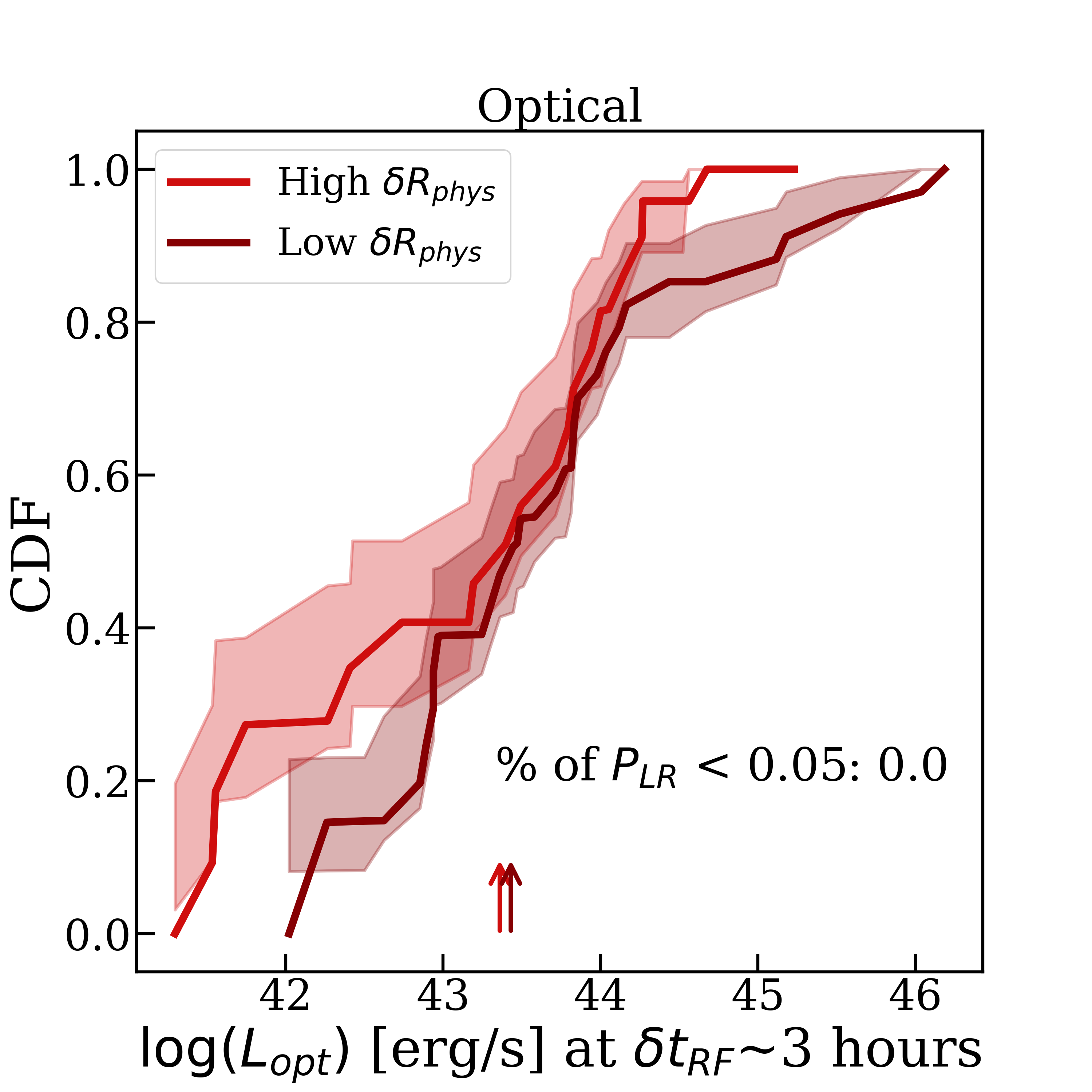}
\includegraphics[width=0.3\textwidth]{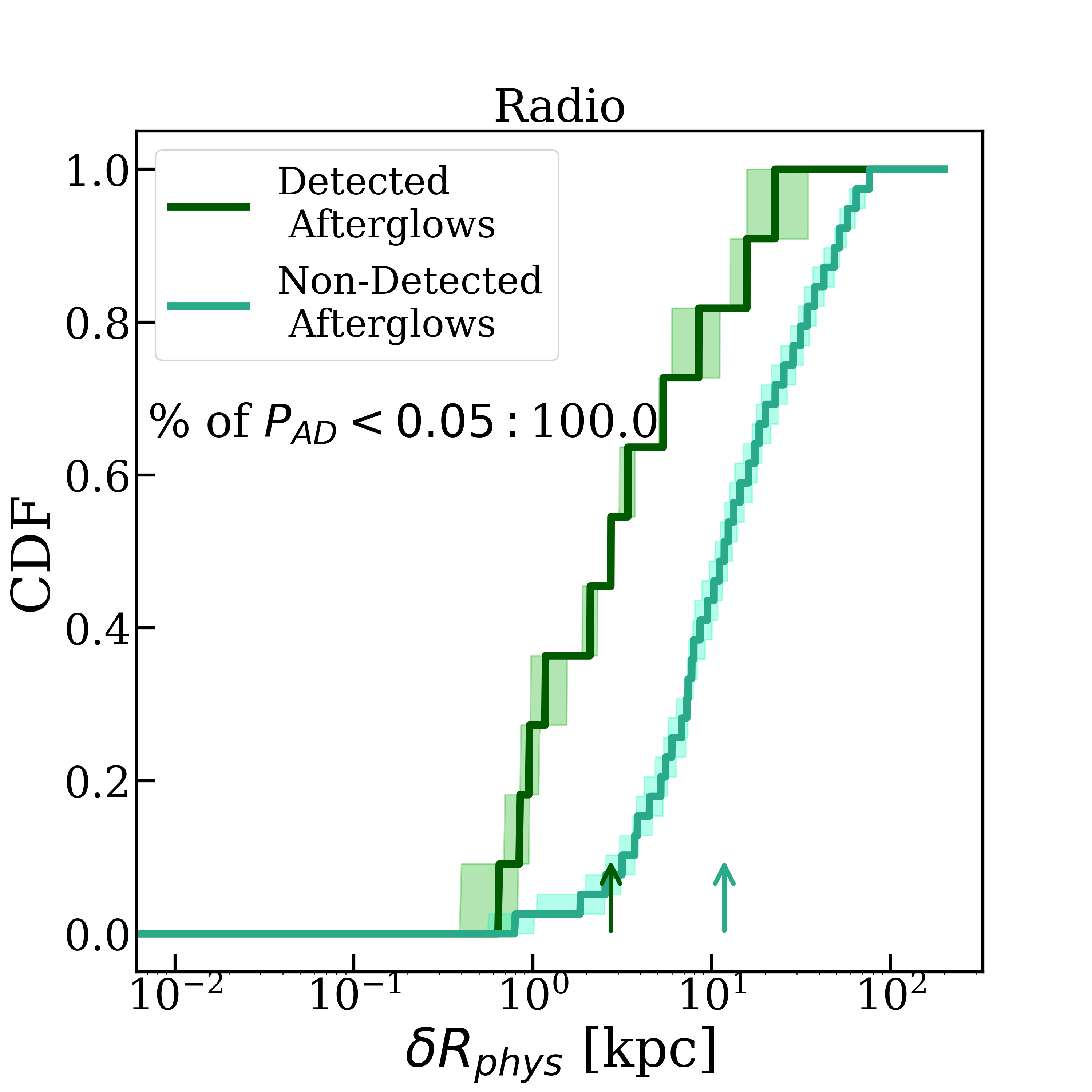}
\includegraphics[width=0.3\textwidth]{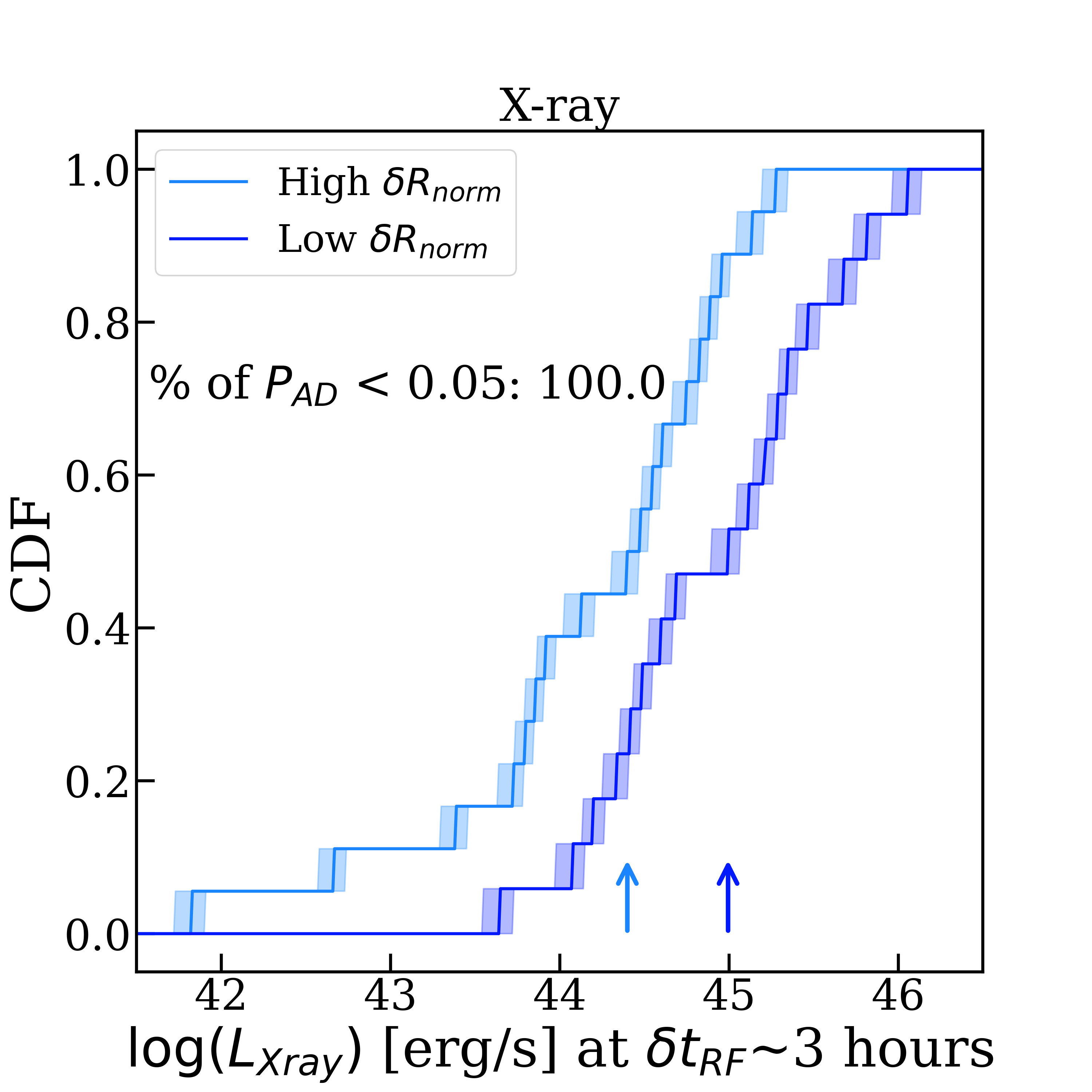}
\includegraphics[width=0.3\textwidth]{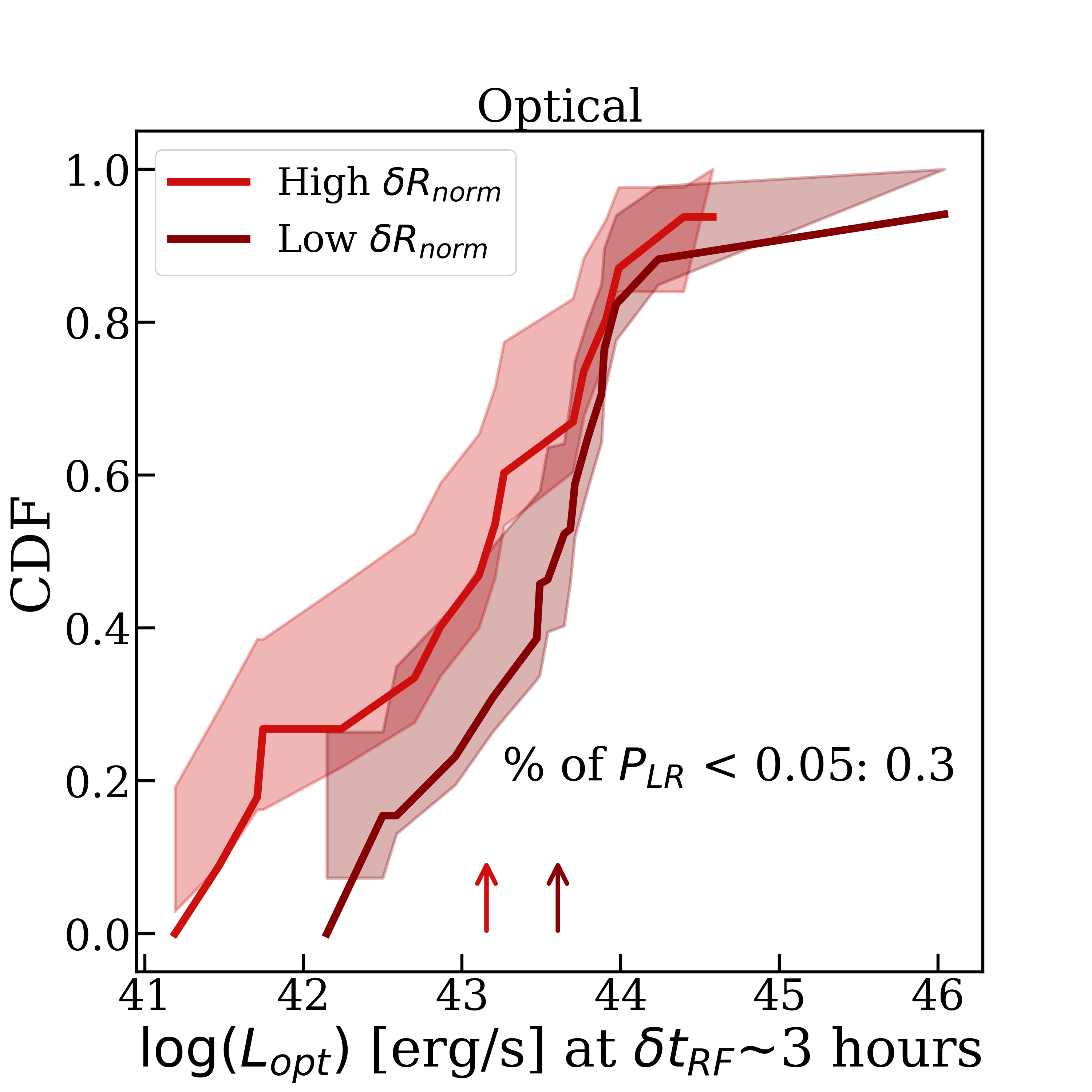}
\includegraphics[width=0.3\textwidth]{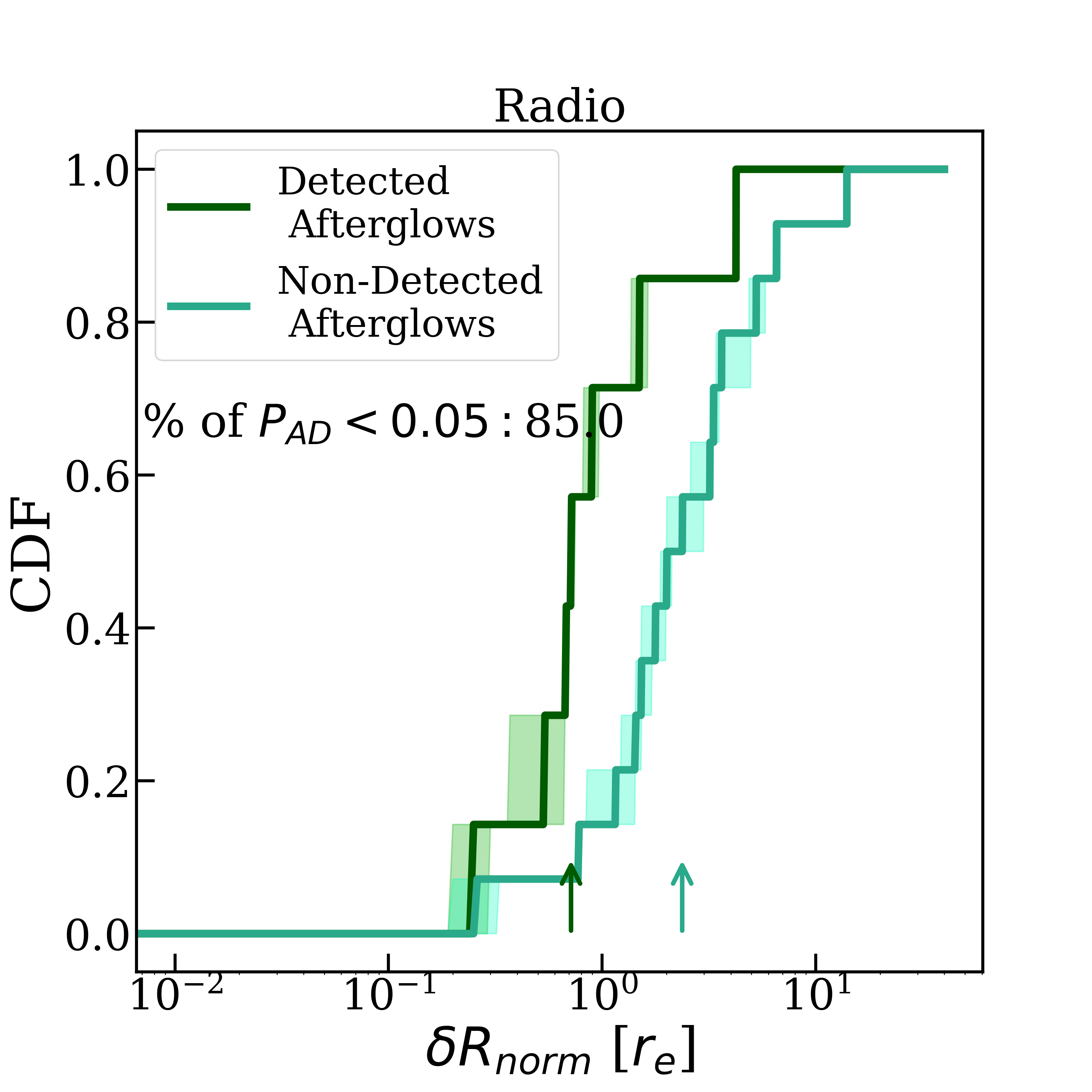}
\caption{\label{fig:offset_cdfs} \textit{Left Column:} CDFs of the X-ray afterglows of short GRBs at $\delta t_c=3$~hr with $\delta R_{\rm phys}>7$~kpc or $\delta R_{\rm norm} > 1.66 \ r_e$ (light blue), and $<7$~kpc or $<1.66 \ r_e$ (dark blue). The median in each distribution is represented by a color-coded arrow denoted from the legend. \textit{Middle Column:} CDFs of the optical afterglows of short GRBs at $\delta t_c=3$~hr with $\delta R_{\rm phys}>7$~kpc or $\delta R_{\rm norm} > 1.66 \ r_e$ (light red) and $<7$~kpc or $<1.66 \ r_e$ (dark red). The median in each distribution is represented by a color-coded arrow denoted from the legend. \textit{Right Column:} CDFs of the $\delta R_{\rm phys}$ or $\delta R_{\rm norm}$ for short GRBs with detected radio afterglows (dark green) versus only upper limits (light green) with arrows pointing at the median. We compare all distributions through AD (X-ray and radio) or logrank (optical) tests. For X-ray optical, and radio afterglows, we show the percentage of tests that reject the null hypothesis. We find that short GRBs with smaller $\delta R_{\rm phys}$ and $\delta R_{\rm norm}$ have brighter X-ray afterglows. In addition, short GRBs with detected radio afterglows tend to have smaller $\delta R_{\rm phys}$.}  
\end{figure*}

For scenario (ii), when there is only a single detection (14 GRBs), we use a single $L_{\rm opt} \propto t^{-1}$ power law to calculate the $L_c$ instead relative to the detection. We calculate $\sigma_{L_C}$ using the median optical afterglow luminosity error, $\sigma_{L_{c}}$ = $0.15 L_c$.  Finally, for scenario (iii), when there are only upper limits (36 GRBs), we simply use the deepest upper limit at $\delta t_c \pm2.5$~hr for a limit $L_c$, as this is the most stringent observational constraint on the GRB afterglow.

\section{Results} \label{sec:results}

Equipped with the afterglow luminosities at a common time ($L_c$) in both the X-ray and optical bands, we next compare these values to host environmental properties. We explore the following properties: galactocentric offsets (physical: $\delta R_{\rm phys}$ and host-normalized: $\delta R_{\rm norm}$), stellar mass ($M_*$), star formation rate (SFR), specific star formation rate (sSFR), and mass-weighted stellar population age ($t_m$). For the radio afterglows, which lack well-sampled light curves from which to extract such luminosities, we instead investigate trends between the detection (or non-detection) of a radio afterglow and any environmental properties.

\subsection{Physical Offsets}
\label{sec:ag_offsets}

\begin{deluxetable}{l c l}
\tabletypesize{\footnotesize}
\tablewidth{1.0\textwidth}
\tablecaption{Split $L_{\textrm{c}}$ Distribution Medians and 68\% Confidence Intervals \label{tab:medians}}
\tablehead{
\colhead{\textbf{Environmental Property}} & 
\colhead{\textbf{$\log(L_{\textrm{c}})$ X-ray}} & 
\colhead{\textbf{$\log(L_{\textrm{c}})$ Optical}} \\
\colhead{} & 
\colhead{[erg/s]} & 
\colhead{[erg/s]}
}
\startdata
$\delta R_\textrm{phys}>7$~kpc & $44.48 \,^{+0.66}_{-1.36}$ & $43.79 \,^{+0.81}_{-1.46}$ \\[2ex]
$\delta R_\textrm{phys}<7$~kpc & $44.68 \,^{+0.89}_{-0.63}$ & $43.55 \,^{+1.16}_{-0.68}$ \\[2ex]
\hline
$\delta R_\textrm{norm}>1.66$ $r_{e}$ & $44.44 \,^{+0.49}_{-0.96}$ & $43.17 \,^{+0.82}_{-1.49}$ \\[2ex]
$\delta R_\textrm{norm}<1.66$ $r_{e}$ & $44.98 \,^{+0.64}_{-0.76}$ &  $43.71 \,^{+0.32}_{-1.07}$ \\[2ex]
\hline
SFR $>1 $ $M_\odot$~yr$^{-1}$ & $44.74 \,^{+0.64}_{-0.91}$ & $43.98 \,^{+0.77}_{-0.65}$ \\[2ex]
SFR $<1 $ $M_\odot$~yr$^{-1}$ & $44.29 \,^{+0.85}_{-1.13}$ & $42.96 \,^{+1.01}_{-1.16}$ \\[2ex]
\hline
sSFR $>10^{-10}$~yr$^{-1}$ & $44.70 \,^{+0.70}_{-0.82}$ & $43.88 \,^{+0.87}_{-0.92}$ \\[2ex]
sSFR $<10^{-10}$~yr$^{-1}$  & $44.19 \,^{+0.80}_{-2.07}$ & $43.22 \,^{+0.87}_{-0.79}$ \\[2ex]
\hline
log($M_*/M_\odot$)$>9.7$ & $44.14 \,^{+1.29}_{-1.31}$ & $43.65 \,^{+1.08}_{-1.23}$ \\[2ex]
log($M_*/M_\odot$)$<9.7$ & $44.62 \,^{+0.62}_{-0.70}$ & $43.81 \,^{+0.86}_{-0.93}$ \\[2ex]
\hline
$t_m >0.8$~Gyr & $44.11 \,^{+0.84}_{-1.32}$ & $43.83 \,^{+0.95}_{-1.45}$ \\[2ex]
$t_m <0.8$~Gyr & $44.75 \,^{+0.76}_{-0.81}$ & $43.57 \,^{+0.87}_{-0.84}$
\enddata
\tablecomments{A table displaying the medians and 68\% confidence intervals of the distribution of luminosities at a common rest-frame time ($L_c$) split by the median environmental property. All medians reported in the X-ray band are calculated from detected afterglows, and the medians reported for optical bands are determined from both detected and non-detected afterglows.}
\end{deluxetable}

We first consider afterglow luminosity and projected galactocentric offsets in physical units ($\delta R_{\rm phys}$). Since the density of the ISM in galaxies generally decreases with increasing radial offset \citep{biegel2012, wjk+2014, ccb+2017, vmr+2024}, and the afterglow brightness scales with circumburst density depending on the observing band \citep{Granot+2002}, we naively expect brighter afterglows at smaller offsets\footnote{We note that we can only measure projected offset, so any possible trend may not be very strong.} (see Section \ref{sec:discussion}). To explore a potential correlation, we divide our population into low-offset ($\delta R_{\rm phys}< 7$~kpc) and high-offset ($\delta R_{\rm phys}> 7$~kpc) events, where $\sim 7$~kpc is the median projected offset for short GRBs \citep{Fong+2022}. For each GRB, we create a probability density function (PDF) for its $L_c$ value, assuming Gaussian uncertainties. We then generate 1000 random draws from each of the $L_c$ PDFs, and build cumulative distribution functions (CDFs) in luminosity for the low- and high-offset populations. We show the X-ray luminosity CDFs in Figure~\ref{fig:offset_cdfs}, and list the luminosity population medians and 68\% confidence intervals in Table~\ref{tab:medians} (along with all other environmental properties studied in this section). As expected, the low-offset events have overall brighter X-ray afterglows than the high-offset events.

For a quantitative comparison, we perform an Anderson Darling (AD) test with the null hypothesis that the distribution of afterglow luminosities at low- and high-offset populations are derived from the same underlying distribution. We compute a $p$-value ($p_{\rm AD}$) for each of the 1000 distributions. If $\geq 50\%$ of the 1000 $p$-values are $p_{\rm AD} < 0.05$, we can reject the null hypothesis that the populations derive from the same underlying distribution.  For the X-ray band, we find that $p_{\rm AD} < 0.05$ for 100\% of the draws, demonstrating a strong rejection of the null hypothesis. This indicates that the low- and high-offset distributions are statistically distinct in their X-ray afterglow luminosities, with low-offset events having brighter afterglows. As mentioned in Section \ref{sec:methods}, we perform the same test at $\delta t_c= 10$ hr to determine if any correlations still hold. We find the difference in afterglow luminosities is still statistically significant at this new common time, validating this observed trend. 

\begin{deluxetable*}{l c c c c c c c c}

\tabletypesize{\footnotesize}
\tablewidth{0pt}

\tablecaption{Afterglow and Environmental Property Distribution Comparisons \label{tab:pvalue_table}}

\tablehead{
\colhead{\textbf{Wavelength}} & 
\colhead{\textbf{Test}} & 
\colhead{\textbf{$\delta R_\textrm{phys}$}} & 
\colhead{\textbf{$\delta R_\textrm{norm}$}} & 
\colhead{\textbf{SFR}} & 
\colhead{\textbf{log(sSFR)}} & 
\colhead{\textbf{log(Stellar Mass)}} & 
\colhead{\textbf{$t_m$}} \\
\colhead{} &
\colhead{} &
\colhead{[kpc]} & 
\colhead{[$r_{e}$]} & 
\colhead{[$M_\odot \ yr^{-1}$]} & 
\colhead{[$yr^{-1}$]} & 
\colhead{[$M_*/M_\odot$]} & 
\colhead{[Gyr]}
}

\startdata
X-ray & \texttt{anderson-ksamp} & \textbf{100\%} & \textbf{100\%} & \textbf{59.9\%} & \textbf{96.7\%} & \textbf{95.9\%} & \textbf{100\%} \\
Optical & \texttt{logrank} & 0.0\% & 0.3\% & \textbf{100\%} & 1.2\% & 0.0\% & 0.0\% \\
Radio & \texttt{anderson-ksamp} & \textbf{100\%} & \textbf{84.9\%} & 0.76\% & 0.08\% & 0.0\% & 0.04\%
\enddata

\tablecomments{The percentage of $p$-values that are $<0.05$, determined either from an AD (X-ray, radio) or logrank (optical) test. For X-ray and optical afterglows, we separate the sample at or near the population median of the given environmental property. For radio afterglows, we display the percentage of $p_{\rm AD}<0.05$, which we use to determine if short GRBs with radio afterglow detections have different environmental properties to those with only upper limits.}

\end{deluxetable*}

In the optical band, we use survival statistics to incorporate the numerous upper limits, and build the CDFs. Specifically, we use a Kaplan–Meier estimator, a non-parametric method that estimates the CDF (survival function). Here, the weight of each upper limit is distributed uniformly with the detections at lower luminosities. Similar to our treatment of X-ray afterglows, we build 1000 distributions with the Kaplan Meier estimator, incorporating each of the random draws of the detected points. Figure~\ref{fig:offset_cdfs} shows that, just like the X-ray afterglows, the low-offset events appear to have brighter optical afterglows, also suggested by the population medians (Table~\ref{tab:medians}). However, there is significant overlap in the high- and low-offset distributions. To quantitatively compare the optical afterglow distributions with upper limits, we employ a Wilcoxon rank-sum test (hereafter, logrank) using the Python \texttt{lifelines}/\texttt{logrank} function \citep{Davidson-Pilon2019}. This statistical test considers the survival of the upper limits, using a Wilcoxon weight and mitigates the influence of upper limits by weighing detected luminosity values more heavily. As with the AD test, we reject the null hypothesis if the majority of the logrank $p$-values are $p_{\rm LR}<0.05$. From the 1000 logrank tests, we find that 0\% of $p_{\rm LR}<0.05$. Thus, despite the apparent differences, we find no statistical support for a difference in the low- and high-offset optical afterglow luminosities. We note that upper limits comprise $\approx 1/3$ of the optical afterglows with offset measurements, and thus their incorporation may make the claim of statistical significance more challenging. To investigate this, we repeat the same analysis with optical afterglow detections only, but recover the same results, signifying that the inclusion of upper limits does not change the results. We also find the same result when performing this test at $\delta t_c= 10$~hr.

For radio afterglows, we take a different approach, dividing the population into those with radio afterglow detections (11 events), and those with only upper limits (39 events). For each GRB in both distributions, we create a probability density function (PDF) for its $\delta R_{\rm phys}$ with its corresponding uncertainty \citep{Fong+2022}. We then generate 1000 random draws from each of the $\delta R_{\rm phys}$ PDFs, and build cumulative distribution functions (CDFs) in $\delta R_{\rm phys}$ for the detected and non-detected radio afterglow populations. We show these CDFs in Figure~\ref{fig:offset_cdfs} and find that those with detections have smaller offsets (medians and 68\% uncertainties of $2.7^{+12.8}_{-1.9}$~kpc for detections, and $11.8^{+27.1}_{-7.5}$~kpc for non-detections). Here, we find that $p_{\rm AD} < 0.05$ for 100\% of the draws, demonstrating a strong rejection of the null hypothesis. This indicates that the detected and non-detected radio afterglows distributions are statistically distinct in their projected physical offsets. Overall, we find that low-offset short GRBs have a higher chance of having a detected radio afterglow. This is consistent with the expected higher ISM densities toward the centers of galaxies, and the strong radio afterglow luminosity dependence on density \citep{Granot+2002}. It also reaffirms the results based on short GRB radio afterglows in \citet{Schroeder+2025}. We place all results of our statistical tests in Table \ref{tab:pvalue_table} for $\delta R_\textrm{phys}$ and all other environmental properties discussed in the following subsections.

\subsection{Host-Normalized Offsets} \label{sec:host_normalized_offset}

We next compare afterglows to the host-normalized offset ($\delta R_{\rm norm}$). As we have shown that afterglows tend to be more luminous at lower physical offset (Section \ref{sec:ag_offsets}), we can extend this expectation to host-normalized offset as well. To investigate this, we divide our population at the median $\delta R_{\rm norm}=1.66 \ r_e$ \citep{Fong+2022} and build 1000 CDFs in the same manner as described in Section \ref{sec:ag_offsets}. As expected, we find that the low-offset events have overall brighter X-ray afterglows than the high-offset ones (Figure~\ref{fig:offset_cdfs}), with 100\% of the $p_{\rm AD} < 0.05$, rejecting the null hypothesis that the two  distributions are drawn from the same underlying distribution. The distributions remain distinct, with statistical backing, when performing the same analysis at $\delta t_c=10$~hr. This fully supports our physical offset results in the X-ray band.

In the optical band, the low-offset events appear to have brighter optical afterglows than the high-offset ones, with slightly less overlap in distributions than the physical offset CDFs (Figure~\ref{fig:offset_cdfs}). However, this apparent difference is not statistically supported: when performing a logrank test between the distributions, we find that 0.3\% of $p_{{\rm LR}}<0.05$, failing to reject the null hypothesis that they are drawn from the same underlying survival distribution. This lack of statistical distinction between optical afterglow luminosity and $\delta R_{\rm norm}$ remains at $\delta t_c=10$ hr, and with detections-only.

In the radio band, of the events with host-normalized offset measurements, 7 events have detections while 14 have upper limits. We find that the median offset for bursts with radio detections is smaller ($0.7 \,^{+1.0}_{-0.4}r_e$), than that of non-detections ($2.0 \,^{+3.6}_{-1.1}r_e$). This difference is statistically significant (85\% of $p$-values have $p_{\rm AD} \leq 0.05$), with radio afterglow events being more detectable at lower host-normalized offsets.

Thus, overall we find that short GRBs have more luminous afterglows at smaller offsets across all wavelengths than those at larger offsets. However, we find that the difference is only statistically significant in the X-ray and radio bands.

\subsection{Star-Formation Rate and Specific Star-Formation Rate} \label{sec:sfr}

\begin{figure*}[t]
\centering
\includegraphics[width=0.3\textwidth]{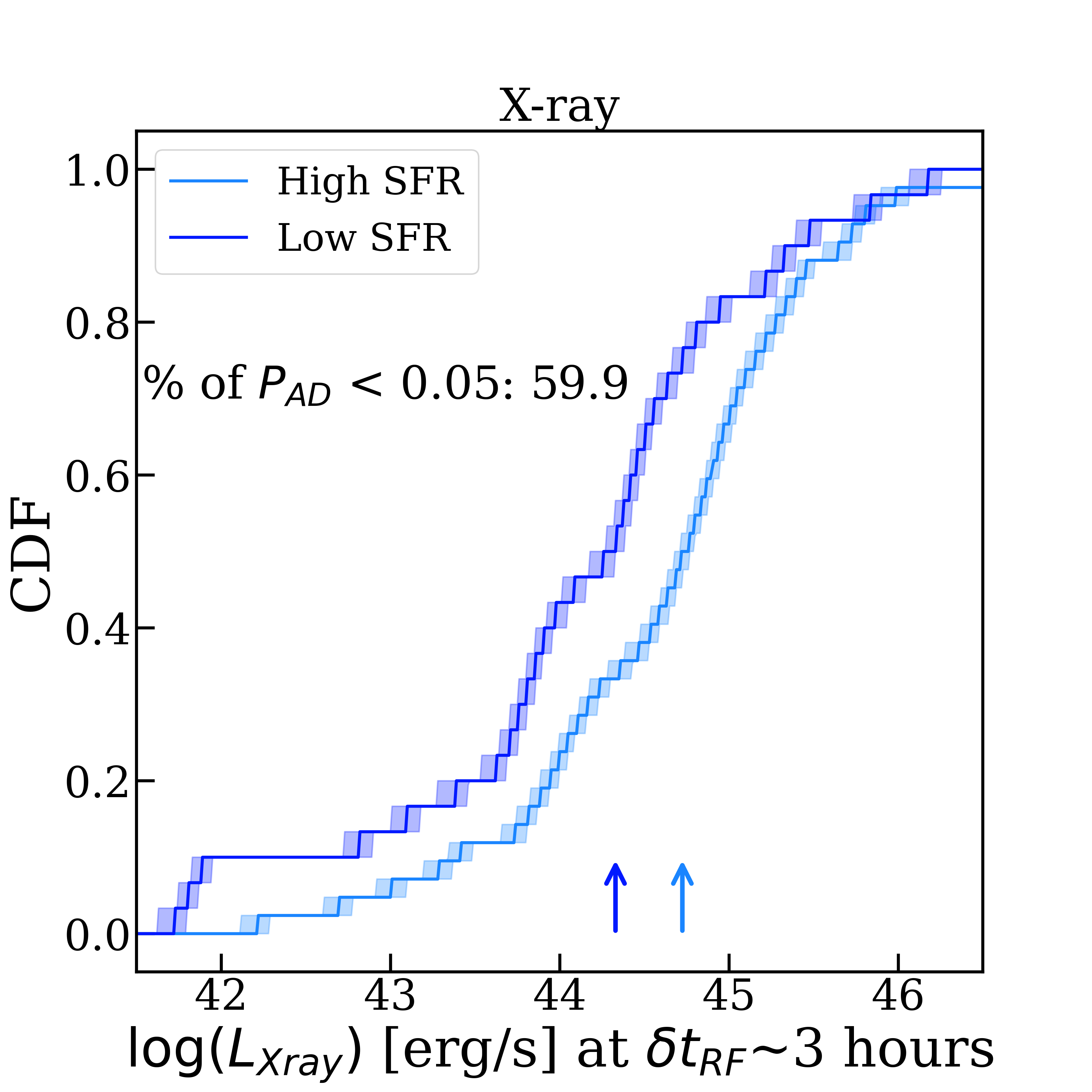}
\includegraphics[width=0.3\textwidth]{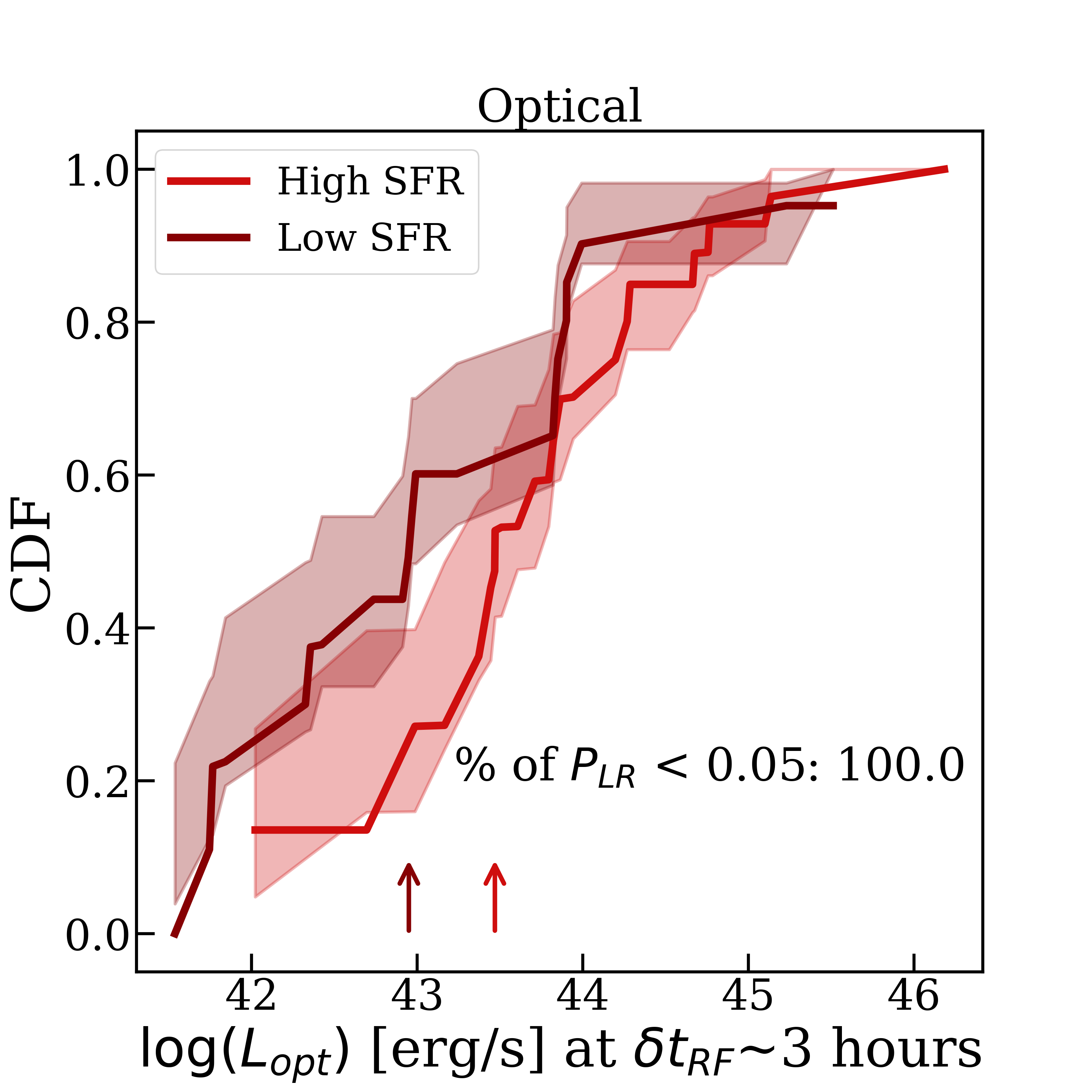}
\includegraphics[width=0.3\textwidth]{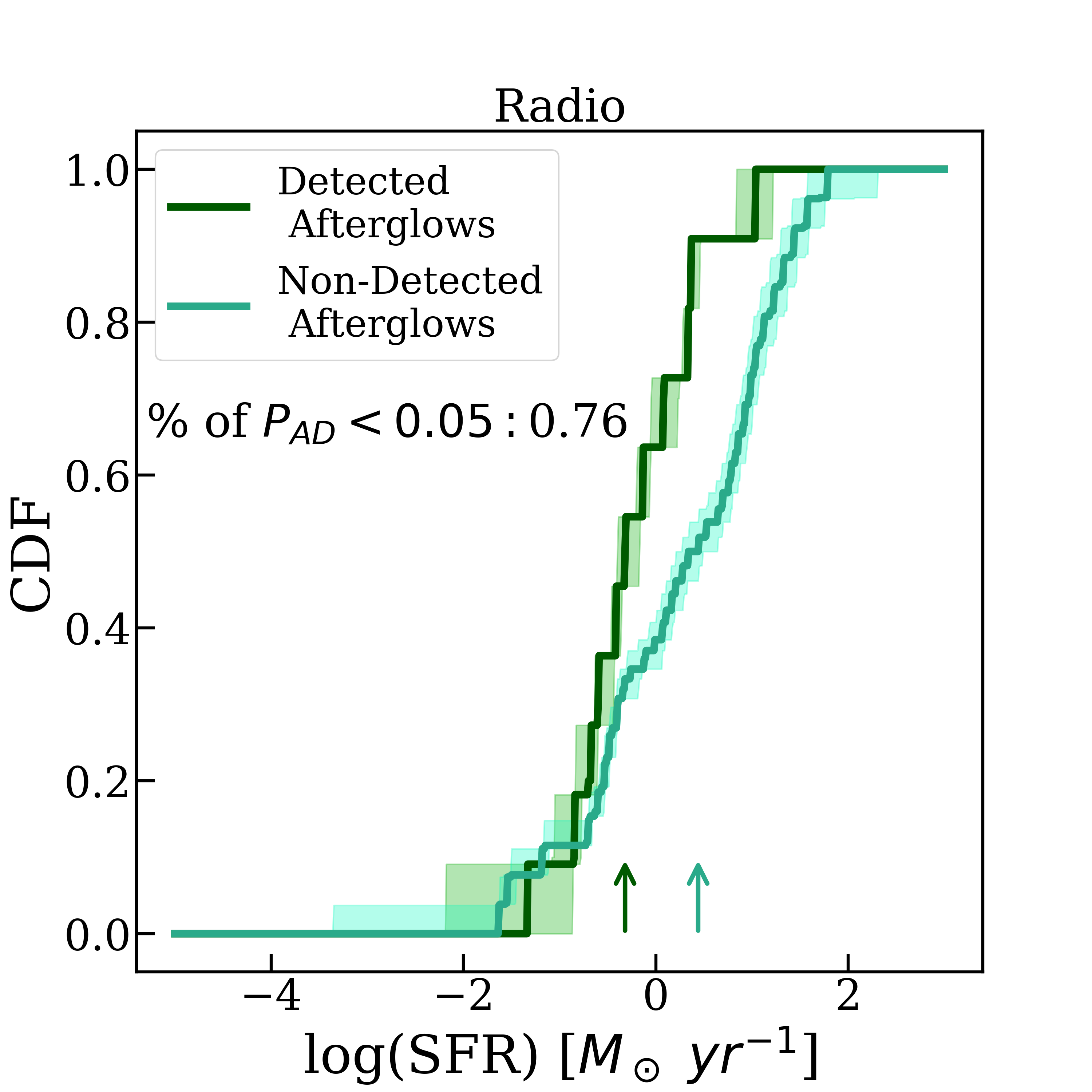}
\includegraphics[width=0.3\textwidth]{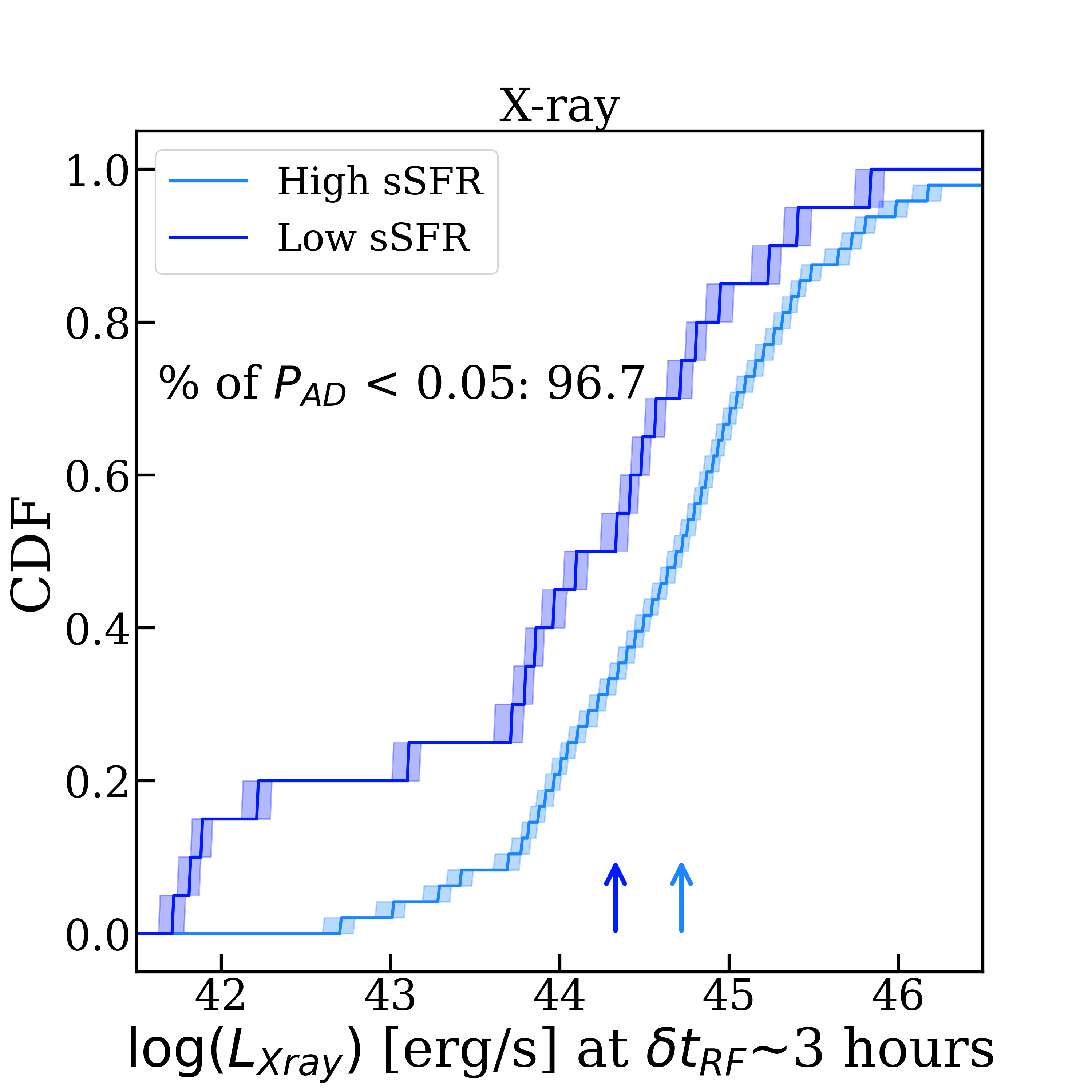}
\includegraphics[width=0.3\textwidth]{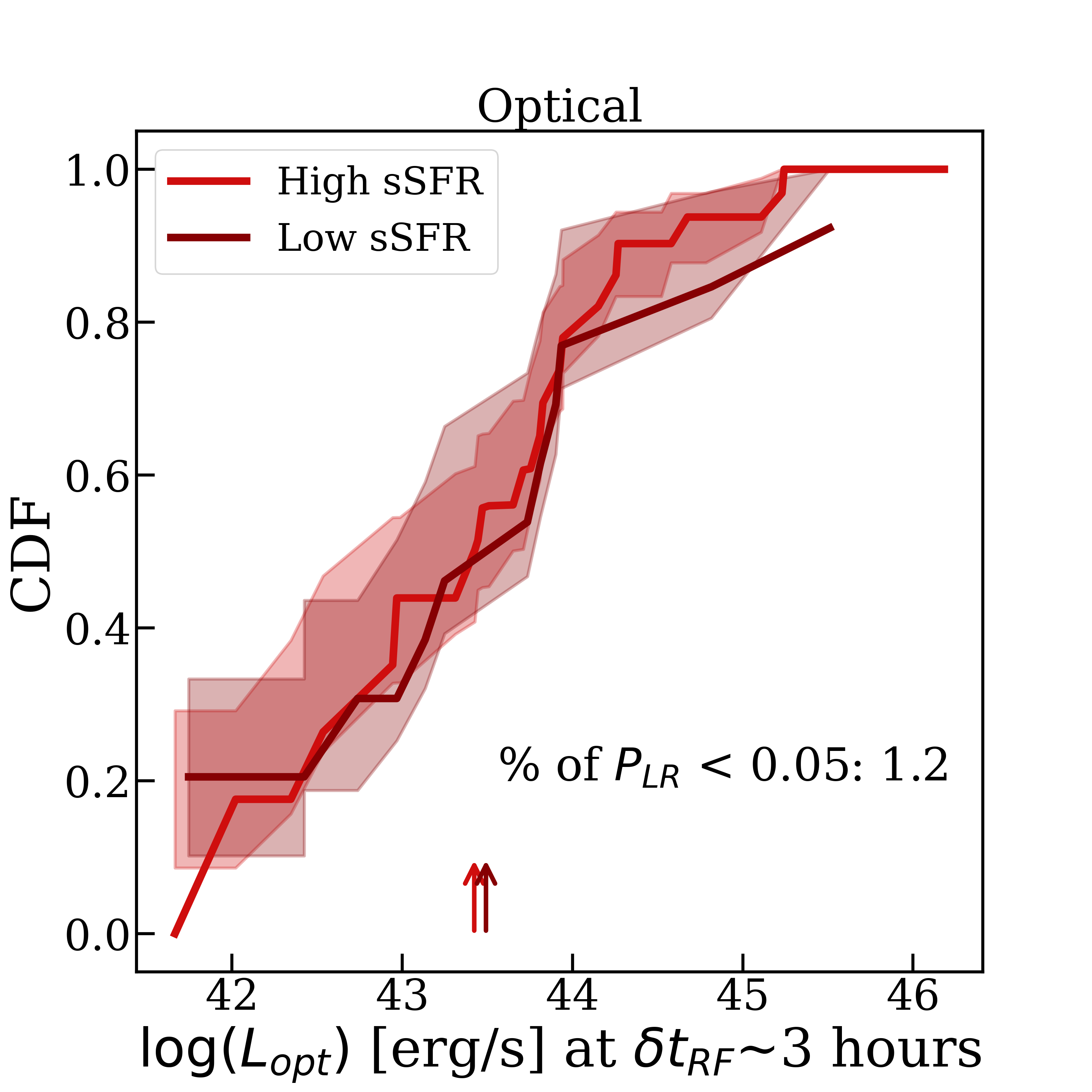}
\includegraphics[width=0.3\textwidth]{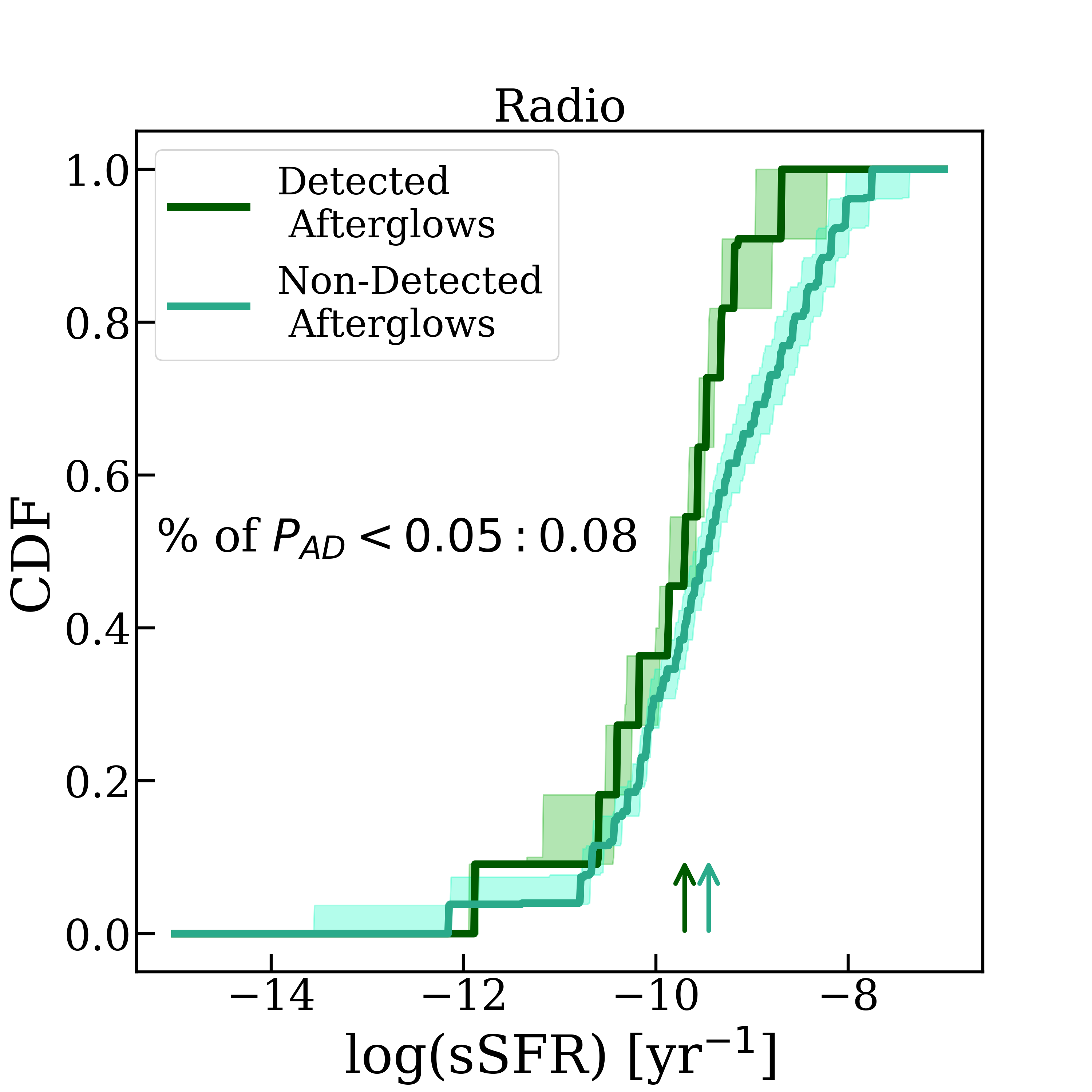}
\caption{\label{fig:SFR_cdfs} The same as Figure \ref{fig:offset_cdfs}, except for SFR and sSFR. For the X-ray and optical afterglow CDFs, the lighter lines represent the short GRBs with SFR $> 1 \ M_\odot$~yr$^{-1}$ or log(sSFR)$>-10$~yr$^{-1}$, and the darker lines show the short GRBs with SFR $< 1 \ M_\odot$~yr$^{-1}$ or log(sSFR)$<-10$~yr$^{-1}$. We find that short GRB X-ray and optical afterglows are more luminous in host galaxies with higher SFR, and X-ray afterglows are additionally brighter in hosts with higher sSFR. In the radio band, it appears that non-detections occur more often in more actively star-forming galaxies but do not pass the threshold for statistical significance.}
\end{figure*}

We next compare afterglows to host galaxy star-formation rates (SFR) and specific SFR (sSFR, defined as SFR/$M_*$). In general, we expect hosts with larger SFR or sSFR values to have higher average ISM densities, leading to more luminous afterglows. We divide our sample into low SFR ($<1\,M_\odot$~yr$^{-1}$) and high SFR ($>1\,M_\odot$~yr$^{-1}$) hosts. Separately, we divide the population into low sSFR ($<10^{-10}$~yr$^{-1}$) and high sSFR ($>10^{-10}$~yr$^{-1}$) distributions\footnote{As sSFR is defined as SFR/$M_*$, higher sSFR is derived from either a host galaxy with higher SFR or with smaller $M_*$, or a combination of both and vice versa.}. Both of the SFR and sSFR thresholds were chosen to roughly align with the median values for the population \citep{Nugent+2022}.

In the X-ray band, we show the distribution of CDFs in Figure~\ref{fig:SFR_cdfs}. The events in high SFR or sSFR hosts appear to have brighter X-ray afterglows than the low SFR or sSFR distributions. For SFR, when comparing the two distributions using an AD test, we find that 60\% of $p_{\rm AD}<0.05$. Thus, we reject the null hypothesis that the two distributions are drawn from the same underlying one. Additionally, we find that the difference is also statistically significant for the sSFR distributions: 100\% of $p_{\rm AD}< 0.05$, and thus we reject the null hypothesis that short GRBs in hosts with high and low sSFR follow the same afterglow luminosity distribution. In other words, we find that short GRB X-ray afterglows tend to be brighter in host galaxies with higher SFR or sSFR. Although we find a correlation between short GRB X-ray afterglow and SFR and sSFR at $\delta t_c=3$ hr, this does not hold at $\delta t_c=10$ hr, likely due to the decreased sample size at 10~hr.

For optical afterglows, Figure~\ref{fig:SFR_cdfs} shows that hosts with higher SFR overall have brighter optical afterglows compared to those with lower SFR. However, the higher and lower sSFR distributions are not as distinct as both distributions have larger overlap. Using a logrank test to compare the SFR distributions with upper limits, we find that 100\% $p_{\rm LR}<0.05$, rejecting the null hypothesis that the two distributions are drawn from the same underlying one. However, we find that the difference {\it is not} statistically significant for the sSFR distributions, with only 1.2\% of $p_{\rm LR}< 0.05$. Thus, this likely indicates that optical afterglows only have a weak dependence on the amount of active star formation in the host. When we perform the same analysis at $\delta t_c = 10$ hr, we do find statistical significance for SFR, but do not for sSFR.

\begin{figure*}[t]
\centering
\includegraphics[width=0.3\textwidth]{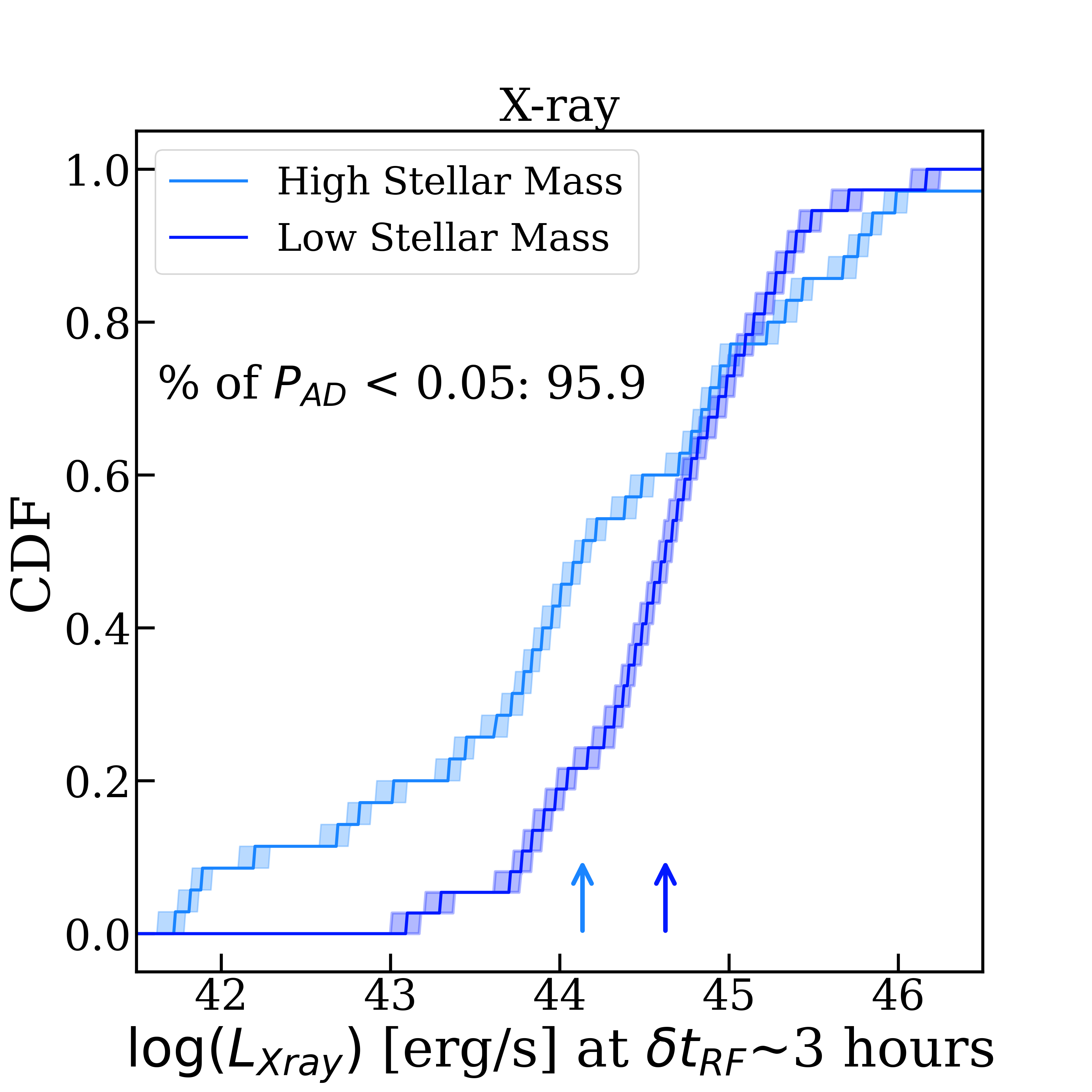}
\includegraphics[width=0.3\textwidth]{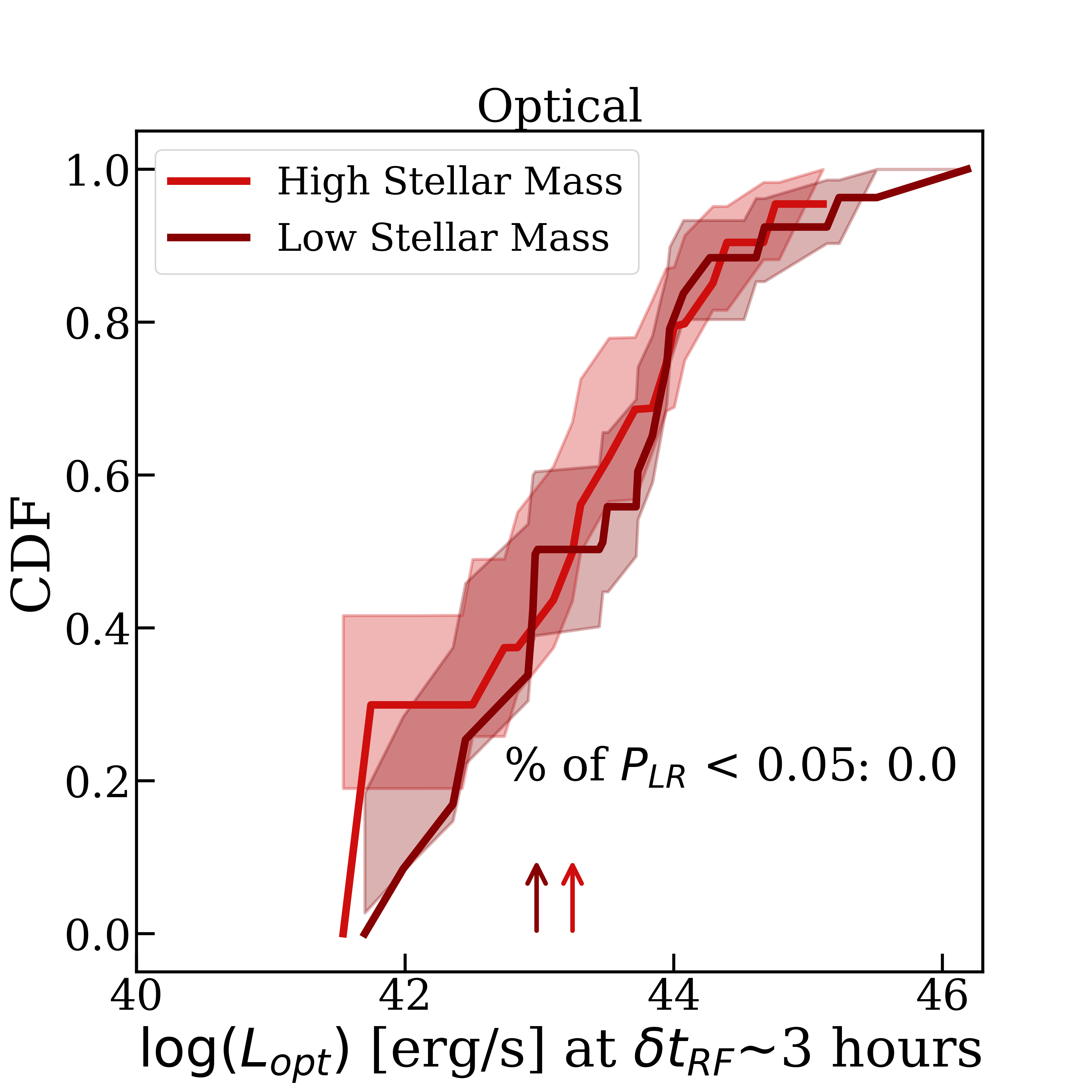}
\includegraphics[width=0.3\textwidth]{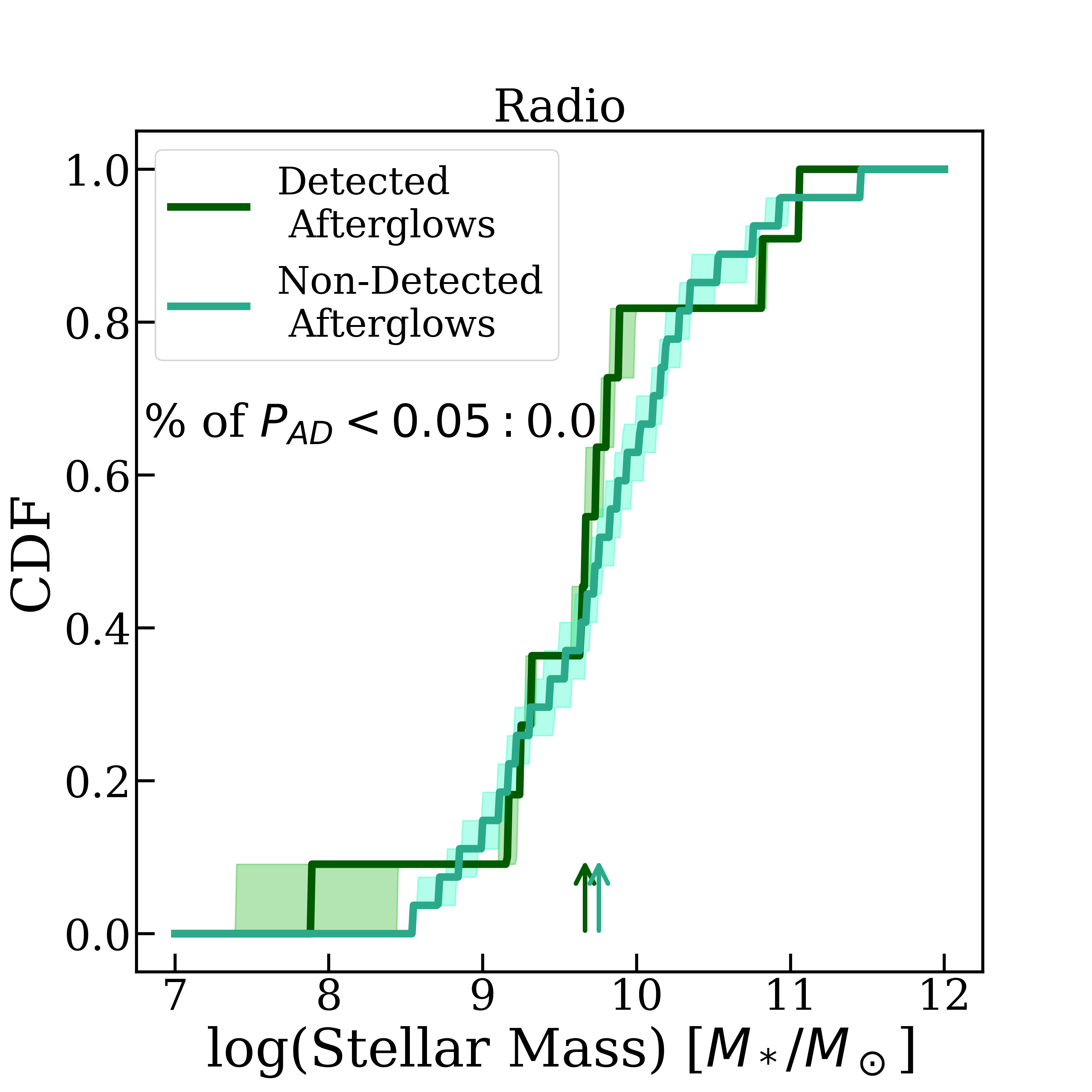}
\caption{\label{fig:stellar_mass_cdfs} The same as Figure \ref{fig:offset_cdfs}, except for stellar mass. For the X-ray and optical afterglow CDFs, the lighter lines represent the short GRBs with $\log(M_*/M_\odot) > 9.7$ and the darker lines show the short GRBs in host galaxies with $\log(M_*/M_\odot) < 9.7$. We find that short GRBs in host galaxies with less stellar mass have brighter X-ray afterglows, although we do not find any correlation between $log(M_*/M_\odot)$ and optical afterglow luminosity or radio afterglow detection.}
\end{figure*}

As in the previous section, we investigate if radio afterglows are correlated with host galaxy SFR or sSFR by computing the relevant distributions for the populations of GRBs with radio afterglows detections (11 events) versus only upper limits (27 events). To accurately probe the uncertainty on SFR and sSFR for the populations of short GRBs with afterglow detections and with only upper limits, we randomly sample 5000 values for each GRB from the SFR and sSFR posteriors provided on the BRIGHT repository \citep{Nugent+2022}. We then build 5000 CDFs from these samples. We find that the median SFR and sSFR for bursts with radio detections is smaller ($0.5 \,^{+1.8}_{-0.4}$ $M_\odot$~yr$^{-1}$ and log(sSFR)=$-9.7 \,^{+0.2}_{-0.1}$~yr$^{-1}$), than those of non-detections ($2.1 \,^{+15.4}_{-1.9}$ $M_\odot$~yr$^{-1}$ and log(sSFR) = $-9.5 \,^{+1.1}_{-0.1}$~yr$^{-1}$). We show the 5000 CDFs of SFRs and sSFRs for the population of GRBs with radio afterglow detections and upper limits in Figure~\ref{fig:SFR_cdfs}. Performing an AD test for the SFR distributions, we find that only 1\% of $p_{\rm{AD}}<0.05$ failing to reject the null hypothesis. Similarly for sSFR, we find that 0.1\% of $p_{\rm AD}< 0.05$, and that the difference {\it is not} statistically significant. Thus, we do not find a clear correlation between star-formation in a host galaxy and detectability of radio afterglows.

Overall, we find that short GRB X-ray and optical afterglows are more luminous in host galaxies that have more active star formation, with X-ray afterglows having a stronger dependence on the amount of star formation in their hosts. However, we do not find a trend in the radio band as the result does not pass the threshold for statistical significance. Our results overall support the picture that the higher expected ISM densities in more star-forming hosts give way to brighter short GRB afterglows.

\subsection{Stellar Mass} \label{sec:stellar_mass}

We next compare afterglows to host galaxy stellar mass ($M_*$). Since quiescent galaxies are typically older and more massive than star-forming galaxies \citep{Blanton+2009, leja2022, tacchella2022}, we can form a reasonable assumption that short GRBs in more massive hosts have less luminous afterglows. This is already hinted at by the fact that we find afterglows in the X-ray (optical) bands are less luminous in low SFR and sSFR (low SFR) galaxies (Section~\ref{sec:sfr}). To investigate this, we divide our sample at the median host stellar mass ($\approx 10^{9.7}$ $M_\odot$; \citealt{Nugent+2022}) into low-mass and high-mass hosts. After building the CDFs in the X-ray band, as expected, the events in hosts with lower stellar masses overall have brighter X-ray afterglows than the events in higher stellar mass hosts (Figure~\ref{fig:stellar_mass_cdfs}). This is statistically supported by our AD test, in which 96\% of the $p_{\rm AD}<0.05$, rejecting the null hypothesis that the two luminosity distributions are drawn from the same underlying one. The same conclusion is drawn when we perform the test at $\delta t_c=10$~hr. This is commensurate with our findings that X-ray afterglows are brighter in hosts with higher SFR and sSFR.

In the optical band, we find that the events that occur in hosts with lower stellar masses also appear to have brighter optical afterglows, although there is overlap at high luminosities (Figure \ref{fig:stellar_mass_cdfs}). This visual difference is not supported statistically, as we find that 0\% of our $p_{\rm{LR}}<0.05$. In contrast to our conclusion with X-ray afterglows, we find that there is no statistical evidence that short GRBs in lower mass hosts have distinct optical afterglow luminosities than those in higher mass hosts despite the visual differences, which stays true when we perform the same tests on detections alone, and for $\delta t_c=10$~hr.

Finally, we investigate any trends for radio afterglows by computing stellar mass distributions for the populations that have radio afterglows detections (11) versus upper limits (27), following the same methods of capturing the uncertainty on the stellar mass distributions as described for SFR and sSFR (Section~\ref{sec:sfr}). For the population of short GRBs with radio afterglow detections, we find a median and 68\% confidence interval of log($M_*/M_\odot$)=$9.7 \,^{+1.1}_{-0.2}$, and a median and 68\% confidence interval of log($M_*/M_\odot$)=$9.8 \,^{+0.5}_{-0.1}$ $M_*/M_\odot$ for the population with upper limits. We show the 5000 CDFs of the stellar masses for the population of GRBs with radio afterglow detections and upper limits in Figure~\ref{fig:stellar_mass_cdfs}. When comparing the distributions with an AD test, we find that 0\% of the $p_{\rm{AD}}<0.05$, failing to reject the null hypothesis that the two distributions are drawn from the same underlying distribution. Thus, we find no support that the detectability of radio afterglows are correlated with stellar mass.

\begin{figure*}[t]
\centering
\includegraphics[width=0.3\textwidth]{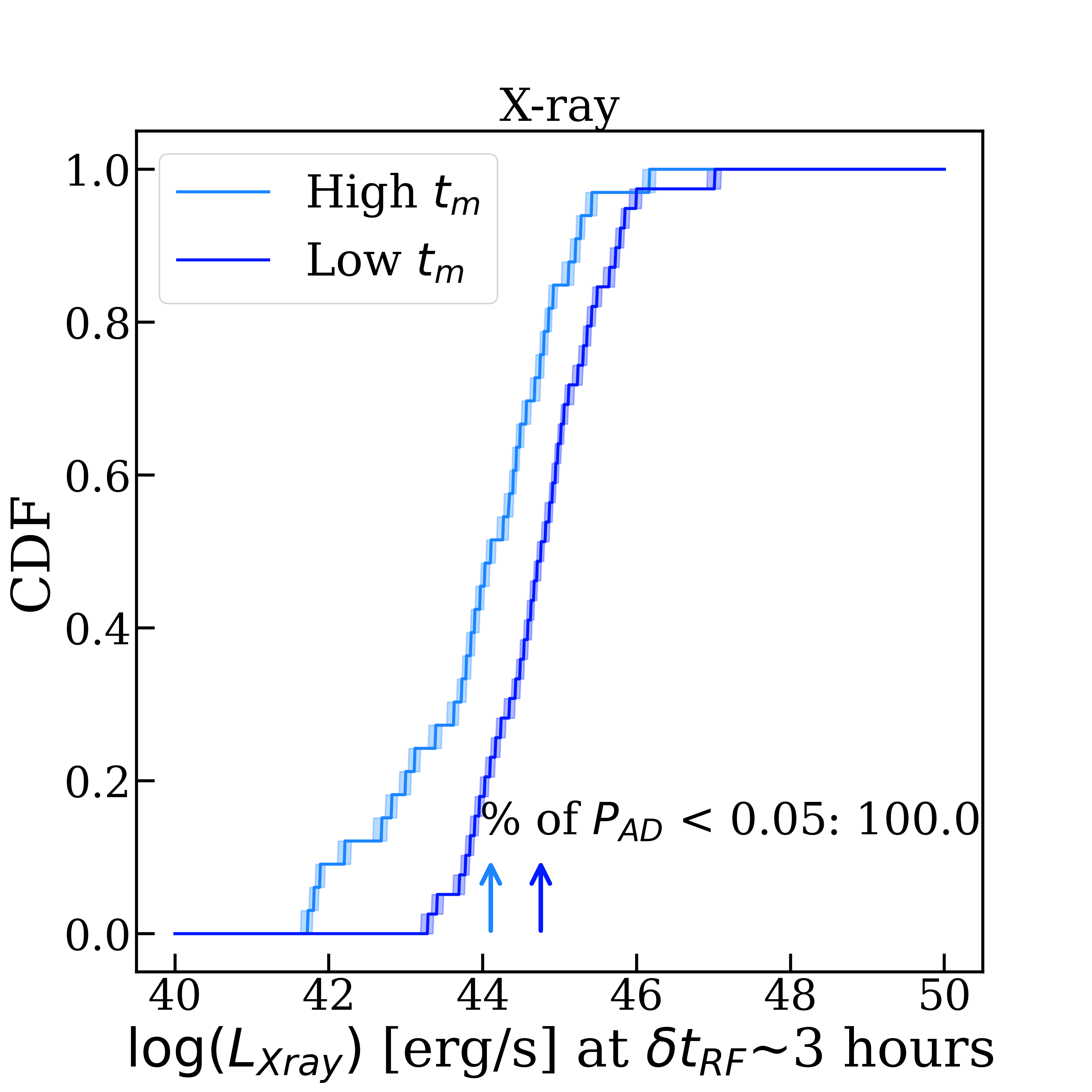}
\includegraphics[width=0.3\textwidth]{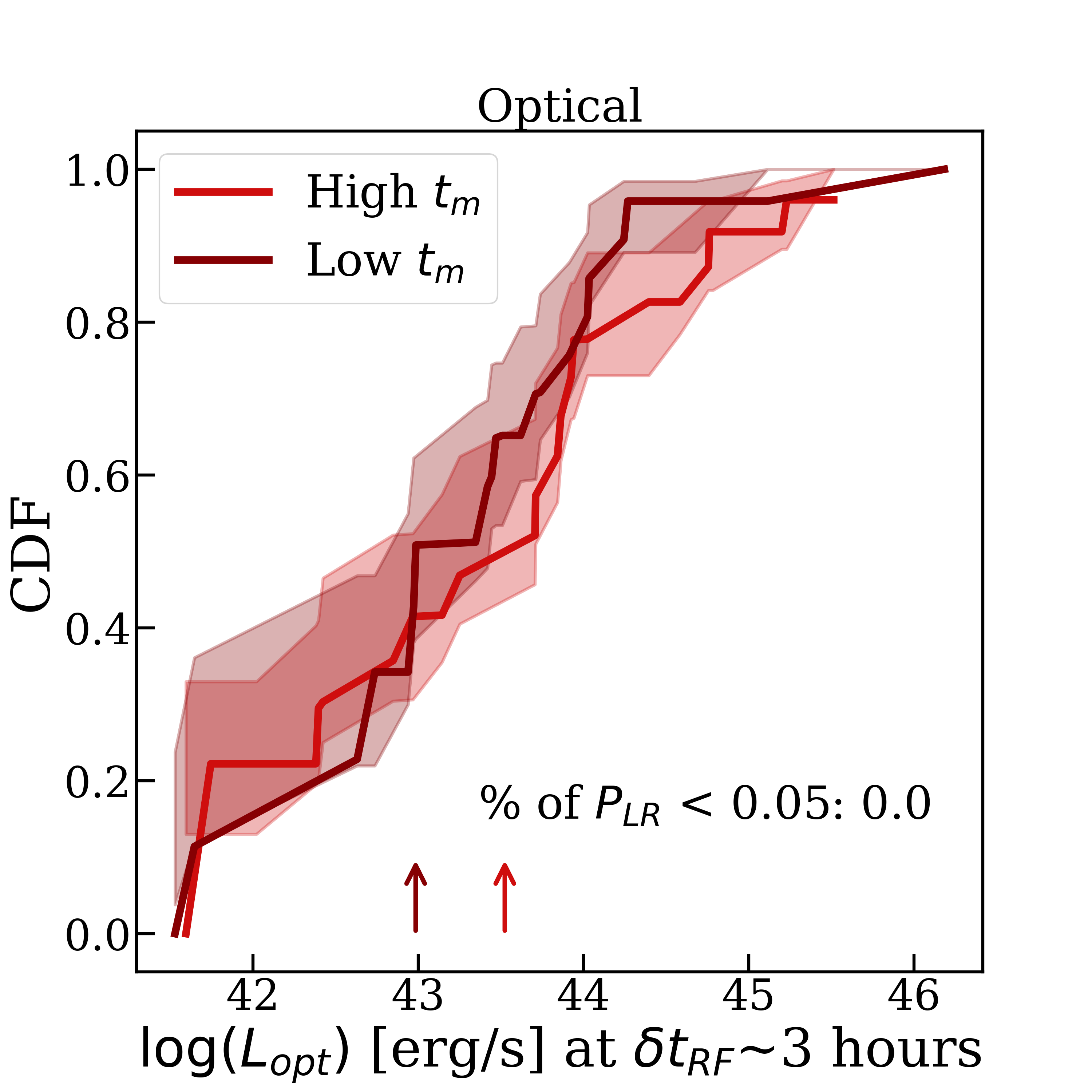}
\includegraphics[width=0.3\textwidth]{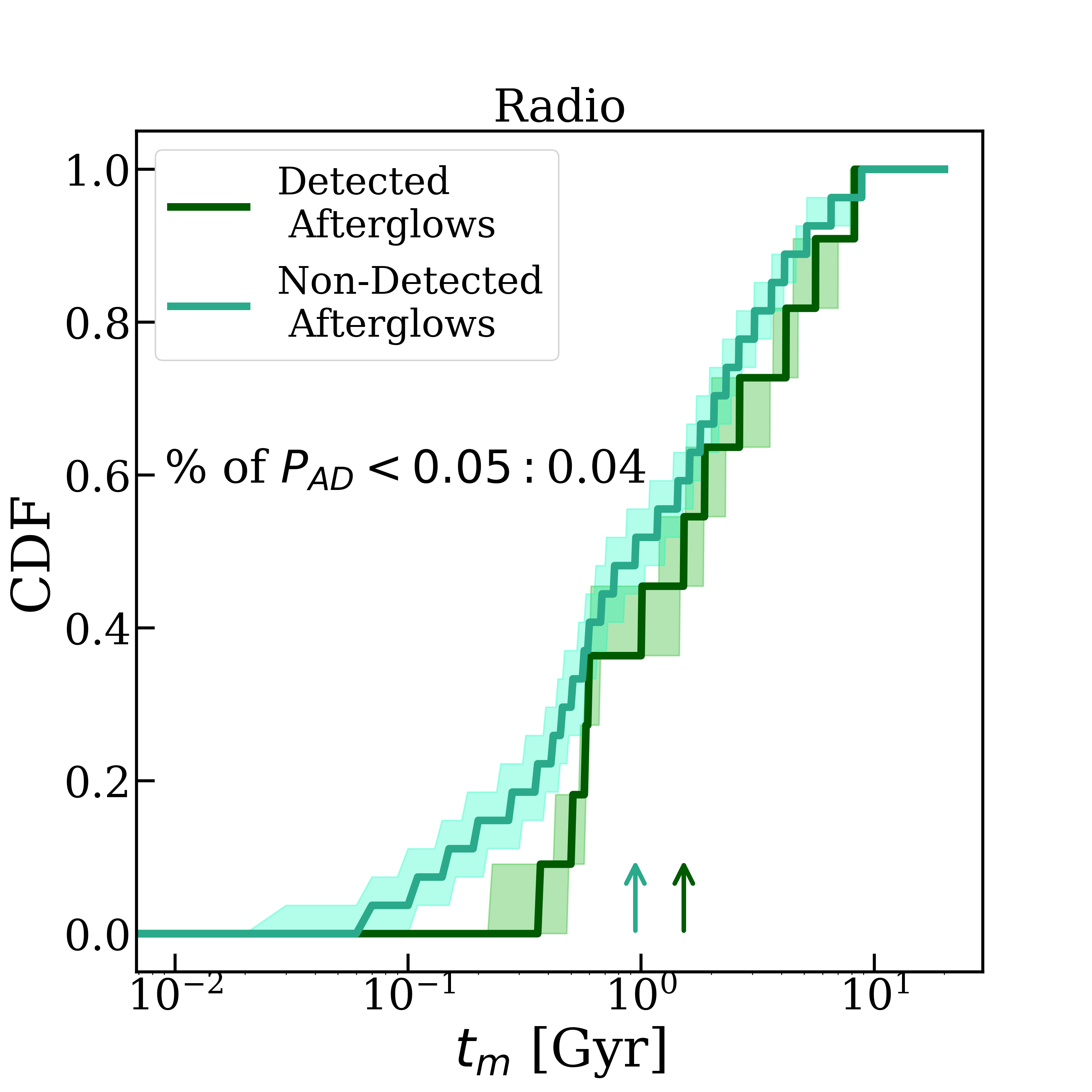}
\caption{\label{fig:t_m} The same as Figure \ref{fig:offset_cdfs}, except for host galaxy stellar population age ($t_m$). For the X-ray and optical afterglow CDFs, the lighter lines represent the short GRBs with $t_m > 0.8$~Gyr and the darker lines show the short GRBs with $t_m < 0.8$~Gyr. We find that short GRBs in host galaxies with lower $t_m$ have brighter X-ray afterglows, although we do not find any correlation between $t_m$ and optical afterglow luminosity or radio afterglow detection.}
\end{figure*}

\subsection{Stellar Population Age} \label{sec:t_m}

Finally, we compare afterglows to host galaxy mass-weighted stellar population age ($t_m$). As previously mentioned, quiescent galaxies are typically older and more massive than star-forming galaxies. Thus, given our previous findings, we expect to find that short GRBs in younger hosts will have more luminous afterglows than those in older hosts. To investigate this, we divide our sample at the median $t_m$ ($\approx 0.8$ Gyr; \citealt{Nugent+2022}) into low $t_m$ (younger) and high $t_m$ (older) hosts. We find that short GRBs in younger hosts have brighter X-ray afterglows than the events in older hosts (Figure~\ref{fig:t_m}). This is statistically supported by our AD test, in which 100\% of the $p_{\rm AD}<0.05$, rejecting the null hypothesis that the two luminosity distributions are drawn from the same underlying one. The same result is found at $\delta t_c=10$~hr.

We do not find the same trend when inspecting the luminosity distributions at high and low $t_m$ in the optical band (Figure \ref{fig:t_m}), as there is significant overlap between the distributions. When comparing the distributions with the logrank test, we furthermore determine that 0\% of our $p_{\rm{LR}}<0.05$ (the same holds true when performing the test for detections only, and also at $\delta t_c=10$~hr for both detections and upper limits). Thus, in contrast to our conclusion with X-ray afterglows, we find that there is no statistical evidence that short GRBs in younger hosts have distinct optical afterglow luminosities than those in older hosts.

Lastly, we compare the distributions of $t_m$ between the population of GRBs with detected radio afterglows and upper limits. We show the CDFs of $t_m$ for these two population in Figure \ref{fig:t_m} and build uncertainties on the CDFs in the same manner as described for SFR and stellar mass (Sections \ref{sec:sfr}-\ref{sec:stellar_mass}). While visually it appears that the population of short GRBs with detected radio afterglows trend toward larger $t_m$ (median and 68\% confidence interval $1.51 \,^{+3.48}_{-0.97}$ Gyr)  than those with only upper limits (median and 68\% confidence interval $0.95 \,^{+2.79}_{-0.69}$), we find that 0\% of the 5000 AD tests performed between the populations result in $p_{\rm{AD}}<0.05$. Thus, we do not find any statistical evidence that detectability of radio afterglows is correlated with $t_m$.

\section{Discussion} \label{sec:discussion}
\subsection{Implications for Local Environments}
\begin{figure}[t]
\centering
\includegraphics[width=0.5\textwidth]{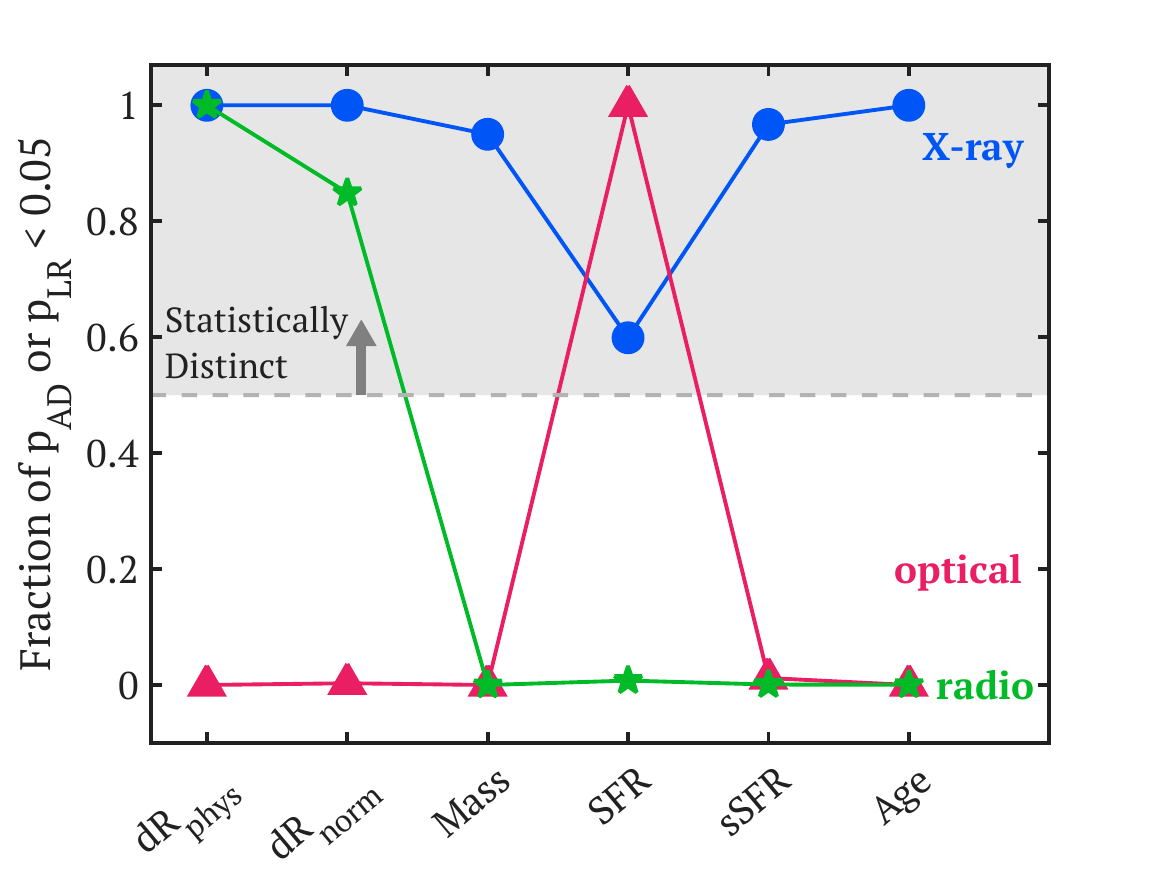}
\caption{\label{fig:pvalues} The fraction of $p_{\rm AD}$ or $p_{\rm LR}$ values that are $<0.05$ when assessing whether distributions of ``high'' and ``low'' (e.g., high-offset and low-offset) environmental properties are drawn from the same underlying distribution. We show the fraction for the X-ray (blue circles), optical (red triangles) and radio (green stars) bands across our tested properties. Fractions above $0.5$ are considered to be support for statistically distinct distributions. Overall, X-ray afterglows align most with expectations, while we only recover statistically significant trends for the radio and optical afterglows in offset and SFR/sSFR, respectively. }
\end{figure}

In this paper, we test whether short GRB afterglow brightness correlates with properties of the kiloparsec-scale environments. The baseline expectation is that brighter afterglows should originate in environments with higher ISM densities - we would typically expect this for GRBs that occur at smaller galactocentric offsets and in host galaxies with higher amounts of active star formation, younger stellar population ages and lower stellar masses. Overall, we find agreement with these expectations, although we do not always find the results to be statistically significant across all wavelengths. To visualize this, in Figure~\ref{fig:pvalues}, we plot the fraction of $p_{\rm AD}$ or $p_{\rm LR}$ values that meet the threshold to refute the null hypothesis ($p<0.05$), where a fraction of $>50\%$ indicates statistically distinct luminosity distributions in that property. We show these values across all tested properties in the X-ray, optical and radio bands.

From Figure~\ref{fig:pvalues}, it is clear that no wavelength has statistically distinct afterglow luminosity distributions across {\it all} properties. We find that X-ray and radio afterglows are brighter or more likely to be detected for GRBs at smaller offsets from their hosts. We also visualize the same trend for optical afterglows, but we cannot recover a statistically significant distinction. We further find that X-ray afterglows are brighter when they occur in hosts with higher amount of star formation (SFR and sSFR), lower stellar mass, and lower stellar population age -- again aligning with all expectations. However, while we also visualize these differences for optical afterglows, the only statistically significant trend we can recover is that they are brighter in hosts with higher SFR; no other tested stellar population property yields a statistical distinction. Finally, we find no other statistically significant trends in the radio band beyond offsets. Taken together, our results demonstrate that the X-ray afterglow luminosity is the most predictable indicator of underlying host properties. Given the visual differences in optical afterglow luminosity distributions, it is possible that increasing the sample size of optical afterglows and/or obtaining deeper afterglow detections and upper limits will lead to a statistical distinction in the future.

We next connect these results to expectations from the local environments. The synchrotron afterglow model provides expectations for how the afterglow brightness maps to local environment (i.e., circumburst density $n_0$) and intrinsic burst properties (energetics, microphysics). Indeed, in the slow-cooling regime, the radio, optical and X-ray bands all share a dependence on circumburst density ($F_{\nu} \propto n_0^{1/2}$ for $\nu_a<\nu<\nu_m$ and $\nu_m<\nu<\nu_c$, where $\nu_a$, $\nu_m$, and $\nu_c$ are the self-absorption, maximum, and cooling break frequencies, respectively; \citealt{Granot+2002}). Thus, all else being equal, afterglows in higher-density environments should be brighter\footnote{The exception is if the X-ray band is above the cooling frequency ($\nu_X>\nu_c$) in which case the afterglow brightness is independent of density \citep{Granot+2002}.} (as shown in the radio band; i.e., \citealt{Schroeder+2025}). The fact that we only recover statistically significant trends with global host properties in the X-ray band could mean that we do not have enough detections to claim statistical significance in the optical or radio bands, that the local and global environmental properties are not strongly linked, or that other factors such as energetics and microphysics play a strong role in afterglow brightness.

\begin{figure*}[t]
\centering
\includegraphics[width=0.495\textwidth]{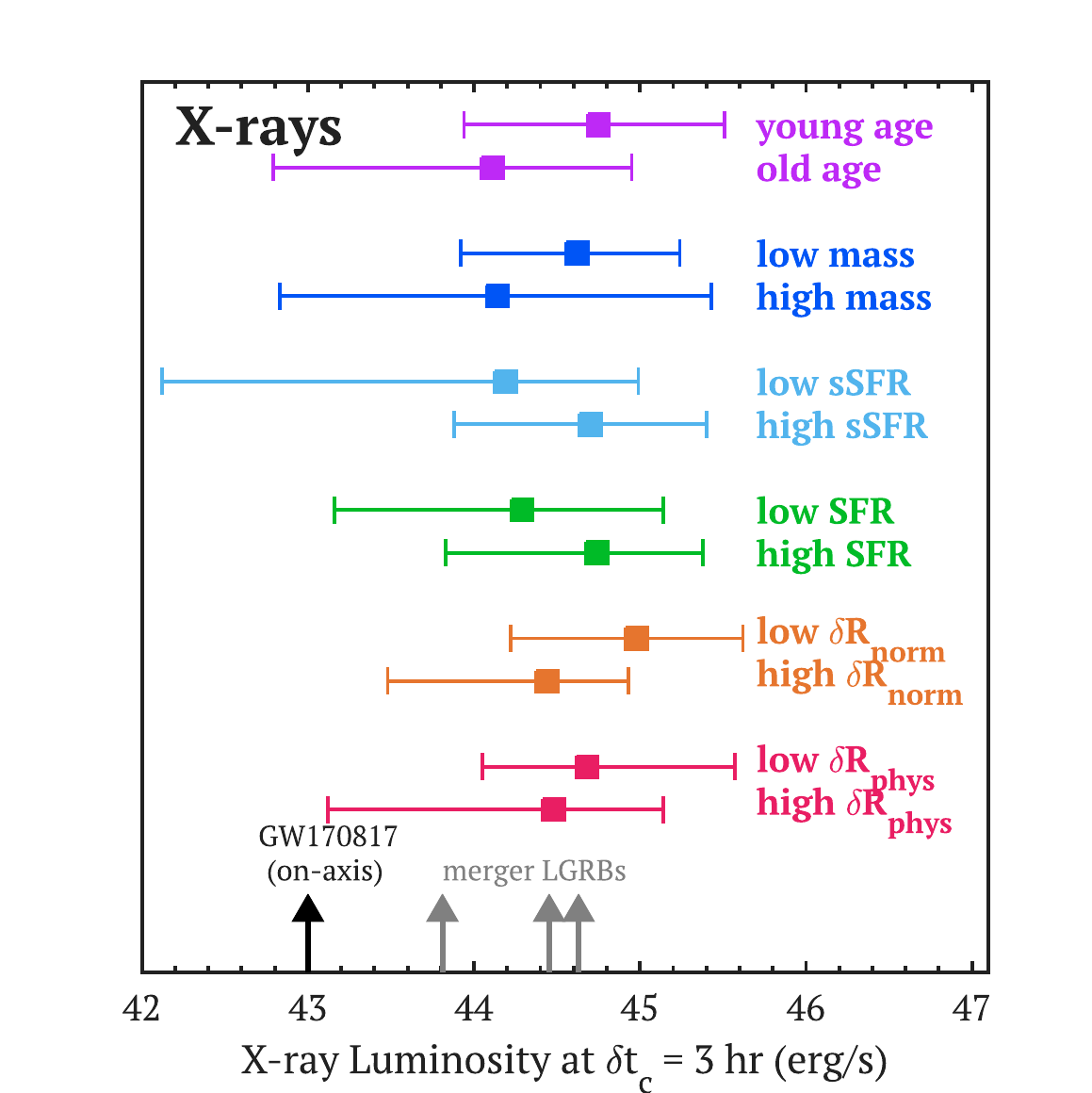}
\includegraphics[width=0.495\textwidth]{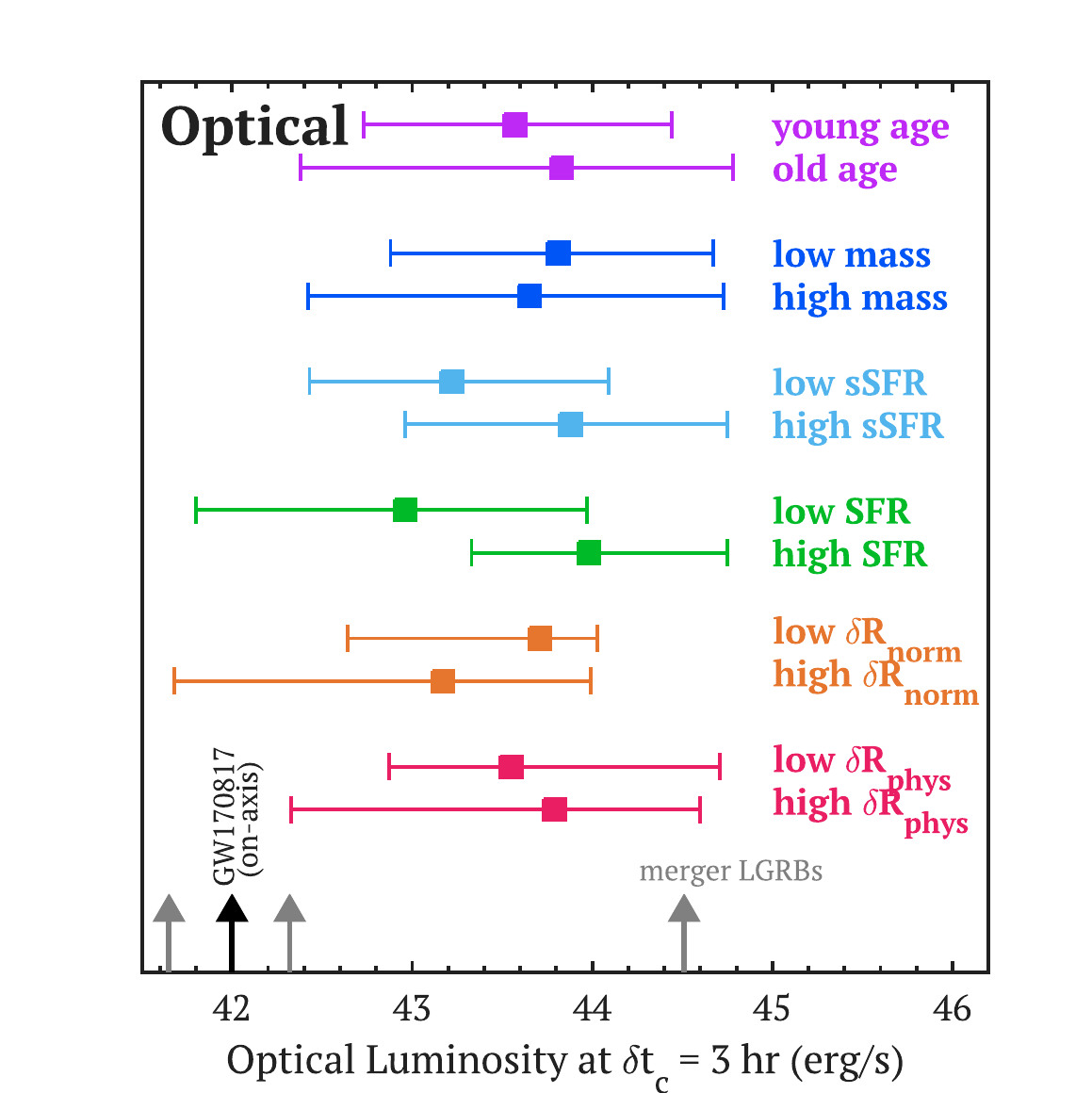}
\caption{\label{fig:lum_bar} Median X-ray (left) and optical (right) afterglow luminosities in which the length of each bar represents the 68\% confidence interval for each population, representing the values in Table~\ref{tab:medians}. Each color represents a different property, split into high- and low-value sub-populations. Overall, we find that short GRB X-ray afterglows are brighter for hosts with younger ages, lower masses, higher SFRs and sSFRs, and for bursts at lower galactocentric offsets. We find that optical afterglows are only brighter in hosts with higher SFR. }
\end{figure*}

\subsection{Merger-Driven Long GRBs and GW Events}

We next turn to the population of long GRBs that likely originate from mergers (``merger-driven LGRBs''; c.f., \citealt{Jin+2015,Yang+2015,Rastinejad+2022,Troja+2022,Yang+2022,Gillanders+2023,Levan+2024,Yang+2024}), and compare their afterglow luminosities to those of short GRBs. For this comparison, we consider GRBs\,060614, 211211A and 230307A as part of the merger-driven LGRB class. We use our analysis to investigate whether their afterglows are systematically brighter than the rest of our sample, as would be expected if they were LGRBs that originated from collapsars (e.g., \citealt{gehrels+2008}). For each environmental property, we compare each GRB's luminosity to the relevant luminosity distribution (i.e., high or low in a given property, as described in Section~\ref{sec:results}). Using the values from Table~\ref{tab:L_crf_results}, we find that all three merger-driven LGRBs fall into the low-SFR and low-mass categories. In terms of offset, GRB\,060614 falls into the low-offset bin, while GRBs\,211211A and 230307A are considered high-offset bursts. Finally, for stellar population age, GRB\,060614 has a young age, while the latter two are in the older age category.  

In Figure~\ref{fig:lum_bar}, we plot the population medians and uncertainties ($68\%$ confidence) in X-ray and optical afterglow luminosities, split by ``high'' and ``low''-values for each tested property. For comparison, we plot the locations of the three merger-driven LGRBs as well as GW170817 (if it was on-axis, see below).  Overall, we find that their X-ray afterglow luminosities are unremarkable compared to the relevant short GRB luminosity distribution (in which $\approx 30-60\%$ of short GRBs are fainter). For optical afterglows, while GRBs\,060614 and 211211A are on the fainter end compared to the population (for which only $\approx 0-33\%$ of afterglows are fainter), we find that GRB\,230307A is very bright in the optical band: at the $\approx 92-96\%$ in brightness, although it is not the brightest. However, this trend is not echoed in the X-rays, as its X-ray afterglow is relatively faint, at the $\approx 11-37\%$ level when compared to the relevant short GRBs. Overall, we do not find clear distinctions between the known merger-driven LGRBs and the traditional short GRB population, lending support that these three events do in fact arise from mergers.

Another highly relevant application of this research is understanding its utility in strategies for follow-up of gravitational-wave (GW) events. Short GRBs are intrinsically related to the local population of GW sources from the association of short GRB 170817A to neutron star (NS) merger GW170817 \citep{Abbott+2017-grb}. Despite the prevalence of follow-up campaigns for subsequent GW-detected binary NS and NS-black hole mergers, there have been no other credible associations of these events to GRBs, afterglow emission, or KNe \citep{andreoni2019,coughlin2019,dobie2019,goldstein2019, gomez2019,hosseinzadeh2019,lundquist2019,ackley2020,andreoni2020,antier2020_1, antier2020_2,garcia2020,gompertz2020,kasliwal2020,morgan2020,pozanenko2020,thakur2020,vieira2020,watson2020,alexander2021,anand2021,becerra2021,bhakta2021,chang2021,deWet2021,dichiara2021,kilpatrick2021,oates2021,ohgami2021,paterson2021,tucker2021,deJaeger2022,dobie2022,rpf+2022,ahumada2024}. However, the lack of detectable electromagnetic counterparts may not be surprising if most other GW-NS mergers are similar to GW170817. The host of GW170817 (NGC4993) has been characterized as massive ($\approx 10^{10.6}M_\odot$), quiescent (SFR $\approx 0.02 M_\odot$~yr$^{-1}$), and old ($\approx10.4$~Gyr; \citealt{Blanchard+2017, llt+2017, pht+17, Pan2017, kilpatrick2022}). \citet{Nugent+2022} places NGC4993 in the top 84\% of stellar masses for short GRB hosts, and determines that it has the lowest SFR (and sSFR) and highest stellar population age compared to all other short GRB hosts (c.f., Figure~\ref{fig:lum_bar}). Although GW170817 has a small galactocentric offset ($\approx2$~kpc), which would generally correlate with a brighter afterglow from our analysis, the rest of its environmental properties indicate that GW170817/GRB170817A should have had a faint afterglow compared to most short GRBs. 

We next more fully test how the afterglow of GRB\,170817A compares to expectations from its host properties given our analysis. First, for direct comparison to our short GRB population, we approximate the X-ray and optical afterglow luminosities of GRB\,170817A had it been {\it on-axis}, since its true observing angle was $\sim 20^{\circ}$ \citep{margutti2017, lazzati2018, lyman2018, resmi2018, troja2018, hotokezaka2019, lamb2019, ghirlanda2019, Salafia2019, mc21}. Using the \citet{wm2019} on-axis model, we find that X-ray and optical afterglow luminosities of $L_X \approx 10^{43}$~erg~s$^{-1}$ and $L_{\rm opt} \approx 10^{42}$~erg~s$^{-1}$ at $\delta t_c=3$~hr. These luminosities are perfectly in alignment with our expectations of a GRB in a quiescent, old and massive environment. The X-ray and optical afterglows are on the faint end when compared to the relevant luminosity distributions (low SFR and sSFR, high stellar mass), with only $\approx 13-29\%$ of afterglows that are fainter (Figure~\ref{fig:lum_bar}). Meanwhile, we note that no events in the low stellar mass category ($<10^{9.7} M_\odot$) have such a low X-ray afterglow luminosity, and only 2-5\% of GRBs with SFR $> 1 \ M_\odot$~yr$^{-1}$ and sSFR $> 10^{-10}$~yr$^{-1}$ have similarly low X-ray afterglow luminosities. We find similar results in the optical band:
no GRBs in high SFR hosts have such a low optical afterglow luminosity and only 6-16\% of GRBs with low stellar mass and high sSFR have lower optical afterglow luminosities.

Thus, if most other local Universe GW-mergers occur in similar environments to GW170817, we expect that their afterglow luminosities will be faint and difficult to detect, especially if the event is more offset from its host (as GW170817 has a low galactocentric offset compared to the short GRB population); this will only be exacerbated because most GW mergers will be off-axis. However, given the diversity of short GRB environments, with the majority in less massive, star-forming galaxies than the host of GW170817, we expect their afterglows to be commensurately brighter. Furthermore, as next generation GW detectors will expand merger populations out to higher redshifts, where the fraction of star-forming galaxies increases, the likelihood of detecting (intrinsically) brighter afterglows may increase. Finally, we note that upcoming observing missions, such as the Vera Rubin Observatory \citep{LSST} and the \textit{Roman} Observatory \citep{Roman}, promise to detect transient emission out to $\approx$26~mag, which will greatly help in detecting faint optical afterglow emission that may be a result of the GRB environment.

\section{Conclusion} \label{sec:conclusion}

In this paper, we leverage a large sample of short GRB X-ray, optical and radio afterglow data, combined with and properties of the GRB environment from uniform host galaxy modeling, to investigate correlations between afterglows and hosts. Our tested environmental properties include physical and host normalized galactocentric offsets, host SFR, sSFR, stellar mass, and stellar population age. Our full sample contains 150 short GRBs observed between 2005-2023, and thus comprises the most extensive exploration of GRB afterglows and their environments to-date. To fairly compare all X-ray and optical afterglow luminosities to environmental properties, we formulate a new method to determine a luminosity at a common rest-frame time ($L_c$) of $\delta t_c = 3$~hours, taking advantage of the full lightcurves in these bands. We then compare the X-ray and optical $L_c$ distributions, divided at (or near) the median value of environmental property of interest. As radio afterglow data is sparse for our short GRB population, we focus our analysis in this band to environmental properties between GRBs with detected and non-detected afterglows. We come to the following conclusions:

\begin{itemize}

\item At small galactocentric offsets (physical and host-normalized; $<7$~kpc, $<1.66r_e$), we find that short GRB X-ray afterglows are more luminous, while radio afterglows are more detectable. Optical afterglows for short GRBs at small offsets also appear to be brighter, but the distinction relative to high-offset short GRBs is not statistically significant.

\item Compared to the expectation that afterglows should be brighter in younger galaxies with higher SFRs or sSFRs and lower stellar masses, we find statistical evidence that X-ray afterglows fully agree with these expectations. In contrast, for optical afterglows, the only statistically significant correlation we find is that they are brighter in galaxies with high SFR. Taken together, our results demonstrate that the X-ray afterglow luminosity is the most predictable indicator of underlying host properties.

\item From a comparison of the three merger-driven LGRBs (GRBs\,060614, 211211A and 230307A) to the traditional short GRB population, we find that their afterglow luminosities are unremarkable compared to the short GRB population. This lends support to the idea that these LGRBs derive from mergers, as opposed to collapsars, which we would expect have brighter afterglows.

\item We estimate the afterglow luminosities of GW170817/GRB\,170817A had it been {\it on-axis}, finding that its inferred X-ray and optical afterglows are on the faint end compared to the short GRB population (with $\approx 13-29\%$ of events fainter in the respective host categories). This aligns well with the expectation that less luminous afterglows arise in quiescent, old and massive environments. Given the correlations we have found, if the host environments of future GW-detected mergers are similar, this may translate to lower afterglow luminosities.
\end{itemize}

\noindent We finally note that it is peculiar that there are less correlations between environmental properties and optical afterglows than there are for environmental properties and X-ray afterglows. Indeed, if afterglow luminosity is correlated with ISM density, than we would expect, for example, to find stronger correlations between optical afterglow luminosity and galactocentric offsets. We postulate that this lack of correlations in the optical band is due to instrument sensitivity, as optical afterglows fade so rapidly that they can quickly go below the detection limit of most optical observatories currently used for follow-up. With the onset of operations of the Vera C. Rubin Observatory, and the upcoming planned {\it Nancy Grace Roman} telescope, the promise of discovering the faintest optical and infrared afterglows may be realized. Rubin and {\it Roman} will also enable detection of the faintest galaxies, thus aiding in characterizing the low-mass and high-redshift GRB hosts. Finally, twenty years of NASA's {\it Swift} Observatory operations have provided an unparalleled opportunity to explore the relationship between short GRBs and their host galaxies. We look forward to the next era of short GRB science, and their crucial connection to multi-messenger counterparts.

    \section*{Acknowledgements}
We are forever grateful to the Neil Gehrels {\it Swift} Observatory, its continued mission to observe GRBs and their afterglows after two decades in-flight, and its ability to inspire new generations of time-domain astronomers.

We acknowledge Jillian Rastinejad for informative discussions. C.C. acknowledges support for this project from the Weinberg Summer Research Grant, the Summer Undergraduate Research Grant from Northwestern University’s Office of Undergraduate Research, and the NASA Illinois Space Grant Consortium. The material contained in this document is based upon work supported in part by a National Aeronautics and Space Administration (NASA) grant or cooperative agreement. Any opinions, findings, conclusions, or recommendations expressed in this material are those of the author and do not necessarily reflect the views of NASA. This work was in part supported through a NASA grant awarded to the Illinois/NASA Space Grant Consortium.

The Fong Group at Northwestern acknowledges support by the National Science Foundation under grant Nos. AST-1909358, AST-2206494, AST-2308182, and CAREER grant No. AST-2047919. The Villar Astro Time Lab acknowledges support through the David and Lucile Packard Foundation, the Research Corporation for Scientific Advancement (through a Cottrell Fellowship), the National Science Foundation under AST-2433718, AST-2407922 and AST-2406110, as well as an Aramont Fellowship for Emerging Science Research. W.F. gratefully acknowledges support by the David and Lucile Packard Foundation, the Alfred P. Sloan Foundation, and the Research Corporation for Science Advancement through Cottrell Scholar Award \#28284. 

The Villar Astro Time Lab acknowledges support through the David and Lucile Packard Foundation, the Research Corporation for Scientific Advancement (through a Cottrell Fellowship), the National Science Foundation under AST-2433718, AST-2407922 and AST-2406110, as well as an Aramont Fellowship for Emerging Science Research. This work is supported by the National Science Foundation under Cooperative Agreement PHY-2019786 (the NSF AI Institute for Artificial Intelligence and Fundamental Interactions). 

We acknowledge the use of public data from the \textit{Swift} data archive.

\vspace{5mm}
\facilities{\textit{Swift} (XRT)}

\software{\texttt{Astropy} \citep{astropy:2022},
          \texttt{Scipy},
          \texttt{lifelines} \citep{Davidson-Pilon2019}
          }

\nocite{*}
\bibliography{references}{}
\bibliographystyle{aasjournal}

\appendix 
\restartappendixnumbering

\section{Sample size}
\label{app:tables}
Here, we show a table that list the $\gamma$-ray $T_{90}$ of each short GRB in our sample, along with its bands in which afterglow follow-up was pursued. We also include the references where we collected the afterglow observations.

\startlongtable
\begin{deluxetable*}{cccccl}
\tabletypesize{\scriptsize}
\tablewidth{0pt}
\tablecaption{Wavelength Availability \label{tab:wavelength_availibility}}
\tablehead{
\colhead{GRB} & 
\colhead{$T_{90}$} & 
\colhead{X-ray} & 
\colhead{Optical} & 
\colhead{Radio} &
\colhead{References}
}
\startdata
050202 & 0.27 & N & Y &  Y$^{a}$ & \cite{GRB050202_GCN3018, GRB050202_GCN3007} \\[1ex] 
050509B & 0.024 & Y & Y & Y & \cite{GRB050509b_GCN3393, GRB050509B_GCN3405} \\[1ex] 
050709$^{b}$ & 0.07 & Y & Y & Y & \cite{Fox+2005} \\[1ex] 
050724A$^{b}$ & 98.0 & Y & Y & Y & \cite{Berger+2005, Panaitescu+2006, Malesani+2007} \\[1ex]
050813 & 0.38 & Y & Y & Y & \cite{GRB050813_GCN3791, GRB050813_GCN3815, GRB050813_GCN3787} \\[1ex] 
050906 & 0.26 & N & Y & Y & \cite{GRB050906_GCN3933, Levan+2005} \\[1ex] 
050925 & \nodata & N & N & Y & \cite{GRB050925_GCN4039} \\[1ex] 
051105A & \nodata & N & N & Y & \cite{GRB051105A_GCN4199} \\[1ex] 
051114A & \nodata & N & N &  Y$^{a}$ & \cite{GRB051114_GCN4276} \\[1ex] 
051210 & 1.3 & Y & Y & N & \cite{GRB051210_GCN4331} \\[1ex] 
051221A & 1.4 & Y & Y & Y & \cite{Soderberg+2006} \\ \hline
060121 & 2.0 & Y & Y & N & \cite{Levan+sep2006} \\[1ex] 
060313 & 0.74 & Y & Y &  Y$^{a}$ & \cite{GRB060313_GCN4877, GRB060313_GCN4884} \\[1ex] 
060502B & 0.09 & N & Y & N & \cite{GRB060502A_GCN5068} \\[1ex] 
060614$^{a,b}$ & 102.0 & Y & Y & Y & \cite{Rastinejad+2021} \\[1ex] 
060801 & 0.5 & Y & Y & Y & \cite{GRB060801_GCN5385, GRB060801_GCN5388,GRB060801_GCN5408} \\[1ex] 
061006 & 129.9 & Y & Y & N & \cite{DAvanzo+2009} \\[1ex] 
061201 & 0.8 & Y & Y & N & \cite{GRB061201_GCN5898} \\[1ex] 
061210 & 85.3 & Y & Y &  Y$^{a}$ & \cite{GRB061210_GCN5912, GRB061210_GCN5910} \\[1ex] 
061217 & 0.2 & N & Y & N & \cite{Berger+2007} \\ \hline
070209 & 0.09 & N & Y & N & \cite{GRB070209_GCN6218} \\[1ex] 
070406 & 1.2 & N & Y & N & \cite{GRB070406_GCN6254} \\[1ex] 
070429B & 0.5 & Y & Y &  Y$^{a}$ & \cite{GRB070429B_GCN6367,GRB070429B_GCN6604,GRB070429B_GCN7140} \\[1ex] 
070707 & 1.1 & N & Y & N & \cite{Piranomonte+2008} \\[1ex] 
070714B$^{b}$ & 64.0 & Y & Y &  Y$^{a}$ & \cite{GRB070714B_GCN6685, Graham+2009} \\[1ex] 
070724A & 0.4 & Y & Y &  Y$^{a}$ & \cite{GRB070724_GCN6667, Berger+2009} \\[1ex] 
070729 & 0.9 & Y & Y &  Y$^{a}$ & \cite{GRB070729_GCN6742, Fong+2015} \\[1ex] 
070809$^{b}$ & 1.3 & Y & Y & N & \cite{Fong+2015} \\[1ex] 
070810B & 0.08 & N & Y & N & \cite{Fong+2015} \\[1ex] 
070923 & \nodata & N & N &  Y$^{a}$ & \cite{GRB070923_GCN6831} \\[1ex] 
071112B & 0.3 & N & Y &  Y$^{a}$ & \cite{GRB071112B_GCN7095,Fong+2015} \\[1ex] 
071227 & 142.5 & Y & Y & N & \cite{Fong+2015} \\ \hline
080121 & 0.7 & N & Y & N & \cite{Fong+2015} \\[1ex] 
080123 & 115.0 & Y & Y & N & \cite{GRB080123_GCN7207, Fong+2015} \\[1ex] 
080426 & 1.7 & Y & Y & N & \cite{GRB080426_GCN, Fong+2015} \\[1ex] 
080503$^{b}$ & 170.0 & Y & Y & Y & \cite{GRB080503_GCN7684, GRB080503_GCN7676, Fong+2015} \\[1ex] 
080702A & 0.5 & Y & Y &  Y$^{a}$ & \cite{GRB080702A_GCN7934, GRB080702A_GCN7932, GRB080702A_GCN7929, Fong+2015} \\[1ex] 
080905A & 1.0 & Y & Y & N & \cite{GRB080905A_GCN8208, GRB080905A_GCN8202, Fong+2015} \\[1ex] 
080919 & 0.6 & Y & Y & N & \cite{GRB080919_GCN8277,Fong+2015} \\[1ex] 
081024A & 1.8 & Y & Y &  Y$^{a}$ & \cite{GRB081024A_GCN8419, GRB081024_GCN8406,GRB081024_GCN8402}; \\ 
& & & & & \cite{GRB081024_GCN8400, Fong+2015} \\[1ex]
081024B & 1.8 & N & Y &  Y$^{a}$ & \cite{GRB081024B_GCN8451,Fong+2015} \\[1ex] 
081226A & 0.4 & Y & Y &  Y$^{a}$ & \cite{GRB081226A_GCN8737,GRB081226a_GCN8952, Fong+2015} \\[1ex] 
081226B & 0.4 & N & Y &  Y$^{a}$ & \cite{GRB081226b_GCN8953, Fong+2015} \\ \hline
090305 & 0.54 & N & Y & N & \cite{Fong+2015} \\[1ex] 
090417A & \nodata & N & N & Y & \cite{GRB090417A_GCN9160} \\[1ex] 
090426 & 1.2 & Y & Y & N & \cite{GRB090426_GCN9292,GRB090426_GCN9265, GRB090426_GCN9312}; \\ 
& & & & & \cite{GRB090426_GCN9267,GRB090426_GCN9266, Fong+2015} \\[1ex]
090510 & 5.66 & Y & Y & Y & \cite{GRB090510_GCN9354, GRB090510_GCN9342, GRB090510_GCN9352, Fong+2015} \\[1ex] 
090515 & 0.036 & Y & Y & Y & \cite{GRB090515_GCN9373,Fong+2015} \\[1ex] 
090607 & 2.3 & Y & Y & N & \cite{Fong+2015} \\[1ex] 
090621B & 0.14 & Y & Y & Y & \cite{GRB090621_GCN9555,Fong+2015} \\[1ex] 
090715A & \nodata & N & N & Y & \cite{GRB090715A_GCN9683} \\[1ex] 
090916 & 0.3 & N & Y & N & \cite{Fong+2015} \\[1ex] 
090927A & \nodata & N & N &  Y$^{a}$ & \cite{GRB090927_GCN10021} \\[1ex] 
091109B & 0.3 & Y & Y & N & \cite{Fong+2015} \\[1ex] 
091117 & 0.43 & N & Y &  Y$^{a}$ & \cite{GRB091117_GCN10182, Fong+2015} \\ \hline
100117A & 0.291 & Y & Y & N & \cite{Fong+2015} \\[1ex] 
100206A & 0.12 & Y & Y & N & \cite{Fong+2015} \\[1ex] 
100213A & \nodata & Y & N & N & \cite{evans2007, evans2009} \\[1ex] 
100625A & 0.33 & Y & Y &  Y$^{a}$ & \cite{Fong+2015} \\[1ex] 
100628A & 0.036 & N & Y &  Y$^{a}$ & \cite{Fong+2015} \\[1ex] 
100702A & 0.16 & Y & Y & N & \cite{Fong+2015} \\[1ex] 
101219A & 0.827 & Y & Y & N & \cite{Fong+2015} \\[1ex] 
101224A & \nodata & Y & N & Y & \cite{Fong+2015} \\ \hline
110112A & 0.524 & Y & Y & Y & \cite{Fong+2015} \\[1ex]
110112B & 0.524 & N & Y & Y & \cite{Fong+2015} \\[1ex]  
110420B & 0.08 & N & Y & Y & \cite{Fong+2015} \\[1ex] 
111020A & 0.384 & Y & Y & Y & \cite{Fong+2012,Fong+2015} \\[1ex] 
111117A & 0.47 & Y & Y & Y & \cite{GRB111117A_GCN12568, Margutti+2012} \\[1ex] 
111121A & \nodata & Y & N &  Y$^{a}$ & \cite{Fong+2015} \\ \hline
120229A & 0.22 & N & Y & Y & \cite{Fong+2015} \\[1ex] 
120305A & 0.1 & Y & Y & Y & \cite{Fong+2015} \\[1ex] 
120521A & 0.45 & Y & Y & Y & \cite{GRB120521A_GCN13338,Fong+2015} \\[1ex] 
120804A & 0.81 & Y & Y & Y & \cite{Fong+2015, Berger+2013_march, GRB120804A_GCN13611} \\[1ex] 
120817B & 0.19 & N & Y & N & \cite{Fong+2015} \\[1ex] 
121226A & 1.0 & Y & Y & Y & \cite{Fong+2015} \\ \hline
130313A & 0.26 & N & Y & Y & \cite{Fong+2015} \\[1ex] 
130515A & 0.29 & Y & Y & N & \cite{Fong+2015} \\[1ex] 
130603B$^{b}$ & 0.18 & Y & Y & Y & \cite{Fong+2014,Fong+2015} \\[1ex] 
130716A & 87.7 & Y & Y & Y & \cite{Fong+2015} \\[1ex] 
130822A & 0.04 & Y & Y & Y & \cite{Fong+2015} \\[1ex] 
130912A & 0.28 & Y & Y & Y & \cite{Fong+2015} \\[1ex] 
131004A & 1.54 & Y & Y &  Y$^{a}$ & \cite{Fong+2015} \\[1ex] 
131125A & 0.5 & N & Y & N & \cite{Fong+2015} \\[1ex] 
131126A & 0.3 & N & Y & N & \cite{Fong+2015} \\[1ex] 
131224A & \nodata & N & N & Y & \cite{Fong+2015} \\ \hline
140129B & 1.36 & Y & Y & N & \cite{Fong+2015} \\[1ex] 
140320A & 0.45 & Y & Y & N & \cite{Fong+2015} \\[1ex] 
140402A & 0.03 & N & Y & N & \cite{Fong+2015} \\[1ex] 
140414A & 0.7 & N & Y & N & \cite{Fong+2015} \\[1ex] 
140516A & 0.19 & Y & Y & Y & \cite{Fong+2015} \\[1ex] 
140606A & 0.34 & N & Y & N & \cite{Fong+2015} \\[1ex] 
140619B & 0.5 & N & Y & Y & \cite{Fong+2015} \\[1ex] 
140622A & 0.13 & Y & Y & Y & \cite{Fong+2015} \\[1ex] 
140903A & 0.29 & Y & Y & Y & \cite{Fong+2015} \\[1ex] 
140930B & 0.84 & Y & Y & Y & \cite{Fong+2015} \\[1ex] 
141205A & 1.1 & N & Y & N & \cite{Fong+2015} \\[1ex] 
141212A & 0.3 & Y & Y & Y & \cite{Fong+2015} \\ \hline
150101B$^{b}$ & 0.018 & Y & Y &  Y$^{a}$ & \cite{Fong+2015, Fong+2016} \\[1ex] 
150101A & \nodata & Y & N & Y & \cite{Fong+2015} \\[1ex] 
150120A & 1.2 & Y & Y & Y & \cite{Fong+2015} \\[1ex] 
150423A & 0.22 & Y & Y & N & \cite{GRB150423A_GCN17763,GRB150423A_GCN17742, GRB150423A_GCN17736}; \\
 & & & & & \cite{GRB150423A_GCN17739, GRB150423A_GCN17750, GRB150423A_GCN17729, GRB150423A_GCN17732} \\[1ex]
150424A & 0.5 & Y & Y & Y & \cite{GRB150424A_GCN17762,GRB150424A_GCN17757,GRB150424A_GCN17751, Fong+2021} \\[1ex] 
150728A & 0.8 & Y & Y & N & \cite{GRB150728A_GCN18092, GRB150728A_GCN18097} \\[1ex] 
150831A & 1.15 & Y & Y & N & \cite{GRB150831A_GCN18219, GRB150831A_GCN18218} \\[1ex] 
151229A & 1.44 & Y & Y & N & \cite{GRB151229A_GCN18753, GRB151229A_GCN18761, GRB151229A_GCN18765} \\ \hline
160303A & 5.0 & Y & Y & N & \cite{GRB160303A_GCN19131,GRB160303A_GCN19127,GRB160303A_GCN19134,GRB160303A_GCN19147}; \\ 
 & & & & & \cite{GRB160303A_GCN19144,GRB160303A_GCN19140, GRB160303A_GCN19128, GRB160303A_GCN19142}; \\
 & & & & & \cite{GRB160303A_GCN19130, GRB160303A_GCN19146} \\[1ex]
160408A & 0.32 & Y & Y & N & \cite{GRB160408A_GCN19267, GRB160408A_GCN19270} \\[1ex] 
160410A & 96.0 & Y & Y & N & \cite{GRB160410A_GCN19300,GRB160410A_GCN19275,GRB160410A_GCN19285,GRB160410A_GCN19274}; \\ 
 & & & & & \cite{GRB160410A_GCN19280,GRB160410A_GCN19272, Fernandez+2023} \\[1ex]
160411A & 0.36 & Y & Y & N & \cite{GRB160411A_GCN19291, GRB160411A_GCN19292} \\[1ex] 
160525B & 0.29 & Y & Y & N & \cite{GRB160525B_GCN19465, GRB160525B_GCN19475, O'Conner+2022} \\[1ex] 
160601A & 0.12 & Y & Y & N & \cite{GRB160601A_GCN19480, GRB160601A_GCN19495, GRB160601A_GCN19496} \\[1ex] 
160624A & 0.192 & Y & Y & N & \cite{GRB160624A_GCN19576,GRB160624A_GCN19571} \\[1ex] 
160821B$^{b}$ & 0.48 & Y & Y & Y & \cite{GRB160821B_GCN19839, GRB160821B_GCN19847,Lamb+2019, Fong+2021} \\[1ex] 
160927A & 0.48 & Y & Y & N & \cite{GRB160927A_GCN19963,GRB160927A_GCN19959} \\[1ex] 
161001A & 2.6 & Y & Y & N & \cite{GRB161001A_GCN19975,GRB161001A_GCN19970, GRB161001A_GCN19999} \\[1ex] 
161104A & 0.1 & Y & Y & N & \cite{GRB161104A_GCN20130, GRB161104A_GCN20137, Nugent+2020} \\ \hline
170127B & 0.51 & Y & Y & Y & \cite{GRB170127B_GCN20549, GRB170127B_GCN20554} \\[1ex] 
170325A & \nodata & N & N & Y & \cite{GRB170325A_GCN20958} \\[1ex]
170428A & 0.2 & Y & Y & Y & \cite{GRB170428A_GCN21050, GRB170428A_GCN21055, GRB170428A_GCN21049, GRB170428A_GCN21058}; \\
 & & & & & \cite{GRB170428A_GCN21051,GRB170428A_GCN21048} \\[1ex]
170524A & \nodata & N & N & Y & \cite{GRB170524A_GCN21212} \\[1ex] 
170728A & 1.25 & Y & Y &  Y$^{a}$ & \cite{GRB170728A_GCN21382, GRB170728A_GCN21386, GRB170728A_GCN21465} \\[1ex] 
170728B & 47.7 & Y & Y & Y & \cite{GRB170728B_GCN21441,GRB170728B_GCN21395,GRB170728B_GCN21374,GRB170728B_GCN21388}; \\
 & & & & & \cite{GRB170728B_GCN21399,GRB170728B_GCN21466,GRB170728B_GCN21419} \\[1ex]
170817 & 2.64 & N & Y & N & \cite{Blanchard+2017,GRB170817_GCN22763} \\ \hline
180402A & 0.18 & Y & Y & N & \cite{GRB180402A_GCN22582, GRB180402A_GCN22587} \\[1ex] 
180418A & 1.9 & Y & Y & N & \cite{GRB180418A_GCN22668,GRB180418A_GCN22659,GRB180418A_GCN22648,GRB180418A_GCN22670}; \\ 
 & & & & & \cite{GRB180418A_GCN22660,Rouco+2021,GRB180418A_GCN22662,GRB180418A_GCN22666} \\
 & & & & & \cite{GRB180418A_GCN22665,GRB180418A_GCN22652,GRB180418A_GCN22664} \\[1ex]
180618A & 47.4 & Y & Y & N & \cite{GRB180618A_GCN22792, GRB180618A_GCN22809, GRB180618A_GCN22810} \\[1ex] 
180727A & 1.06 & Y & Y & N & \cite{GBR180727A_GCN23062} \\[1ex] 
180805B & 122.5 & Y & Y & N & \cite{GRB180805B_GCN23090,GRB180805B_GCN23080, GRB180805B_GCN23089} \\[1ex] 
181123B & 0.26 & Y & Y & Y & \cite{GRB181123B_GCN23467,GRB181123B_GCN23439, GRB181123B_GCN23437, GRB181123B_GCN23440}; \\ 
& & & & & \cite{Paterson+2020} \\ \hline
191031D & 0.29 & Y & Y & Y & \cite{GRB191031D_GCN26121,GRB191031D_GCN26147, GRB191031D_GCN26128,GRB191031D_GCN26120}; \\ 
 & & & & & \cite{GRB191031D_GCN26125,GRB191031D_GCN26134} \\ \hline
200219A & 0.5 & Y & Y & Y & \cite{GRB200219A_GCN27245,GRB200219A_GCN27165} \\[1ex] 
200411A & 0.22 & Y & Y & Y & \cite{GRB200411A_GCN27573,GRB200411A_GCN27539, GRB200411A_GCN27542} \\[1ex] 
200522A$^{b}$ & 0.616 & Y & Y & Y & \cite{GRB200522A_GCN27822,GRB200522A_GCN27784,GRB200522A_GCN27783,GRB200522A_GCN27789}; \\ 
 & & & & & \cite{GRB200522A_GCN27792, GRB200522A_GCN27794,Fong+2021} \\[1ex]
200907B & 0.83 & Y & Y & N & \cite{GRB200907B_GCN28400} \\[1ex] 
201006A & 0.49 & Y & Y & Y & \cite{GRB201006A_GCN28572,GRB201006A_GCN28571, GRB201006A_GCN28565, Schroeder+2025} \\[1ex] 
201221D & 0.16 & Y & Y & N & \cite{GRB201221D_GCN29129,GRB201221D_GCN29128,GRB201221D_GCN29148, GRB201221D_GCN29122}; \\ 
 & & & & & \cite{GRB201221D_GCN29117,GRB201221D_GCN29118} \\ \hline
210323A & 1.12 & Y & Y & Y & \cite{GRB210323A_GCN29712,GRB210323A_GCN29717, GRB210323A_GCN29700,GRB210323A_GCN29703}; \\
 & & & & & \cite{GRB210323A_GCN29710,GRB210323A_GCN29720,GRB210323A_GCN29724,Schroeder+2025} \\[1ex]
210726A & 0.38 & Y & Y & Y & \cite{GRB210726A_GCN30534, Schroeder+2024} \\[1ex] 
210919A & 0.16 & Y & Y & Y & \cite{GRB210919A_GCN30856,GRB210919A_GCN30852,GRB210919A_GCN30857,GRB211023B_GCN30992}; \\ 
 & & & & & \cite{GRB210919A_GCN30855} \\[1ex]
211023B & 1.3 & Y & Y & Y & \cite{GRB211023B_GCN31016,GRB211023B_GCN31031,GRB211023B_GCN30976,GRB211023B_GCN30975}; \\ 
 & & & & & \cite{GRB211023B_GCN30975,GRB211023B_GCN31000,GRB211023B_GCN30972,GRB211023B_GCN31002}; \\
 & & & & & \cite{GRB211023B_GCN31107,GRB211023B_GCN31038,GRB2121023B_GCN31021,GRB211023B_GCN30973} \\[1ex]
211106A & 1.75 & Y & Y & Y & \cite{GRB211106A_GCN31063, GRB211106A_GCN31069, Laskar+2022} \\[1ex] 
211211A$^{b}$ & 51.37 & Y & Y & Y & \cite{GRB211211A_GCN31222,GRB211211A_GCN31218,GRB211211A_GCN31217,GRB211211A_GCN31213}; \\ 
 & & & & & \cite{GRB211211A_GCN31227,GRB211211A_GCN31235,GRB211211A_GCN31216,GRB211211A_GCN31221}; \\ 
 & & & & & \cite{GRB211211A_GCN31214,GRB211211A_GCN31203,Rastinejad+2022} \\ \hline
220412B & \nodata & N & Y & N & \cite{GRB220412B_GCN31904,GRB220412B_GCN31892, GRB220412B_GCN31889, GRB220412B_GCN31917} \\[1ex] 
220730A & \nodata & N & Y & N & \cite{GRB220730A_GCN32435} \\[1ex] 
220831A & \nodata & N & N & Y & \cite{GRB220831A_GCN32529} \\[1ex] 
221120A & \nodata & N & Y & N & \cite{GRB221120A_GCN32966, GRB221120A_GCN32958, GRB221120A_GCN32968,GRB221120A_GCN32978}; \\ 
 & & & & & \cite{GRB221120A_GCN32956} \\ \hline
230205A & \nodata & Y & Y & Y & \cite{GRB230205A_GCN33309,GRB230205A_GCN33372, GRB230205A_GCN33289}; \\
 & & & & & \cite{GRB230205A_GCN33286,GRB230205A_GCN33282} \\[1ex]
230217A & \nodata & Y & Y & Y & \cite{GRB230217A_GCN33433, GRB230217A_GCN33365,GRB230217A_GCN33374,GRB230217A_GCN33356}; \\
 & & & & & \cite{GRB230217A_33358,GRB230217A_GCN33370,GRB230217A_GCN33360} \\[1ex]
230228A & \nodata & Y & Y & Y & \cite{GRB230228A_GCN33383,GRB230228A_GCN33392,GRB230228A_GCN33379,GRB230228A_GCN33387}; \\
 & & & & & \cite{GRB230228A_GCN33436,GRB230228A_GCN33381,GRB230228A_GCN33426,GRB230228A_GCN33394} \\[1ex]
230307A$^{b}$ & 33.0 & Y & Y & Y & \cite{GRB230307A_GCN33414,GRB230307A_GCN33412} \\[1ex] 
231117A & 0.668 & Y & Y & Y & \cite{Schroeder+2025} \\ 
\enddata
\tablecomments{The $T_{90}$ and presence of an X-ray, optical, and radio afterglow observation for the 151 short GRBs within our sample. Here, ``Y" denotes an observation (detection or upper limit), ``N" denotes the absence of one. Refer to \cite{evans2007, evans2009, Fong+2015, rfb+2023} for the X-ray afterglow sample. \\ {$^a$}{Shallow radio follow up ($>100 \ \mu$Jy)} \\
{$^b$}{GRBs with possible KNe affecting their optical afterglow. We give the time intervals here for which KN emission may dominate, and avoid these time ranges in our analysis: GRB\,050709: $\delta t \approx 1-4$~days \citep{Gompertz+2018, Rossi+2020, Rastinejad+2021}; GRB\,050724A: $\delta t \approx 1.45 - 3.46$~days \citep{Rossi+2020, Rastinejad+2021}; GRB\,060614: $\delta t \approx 3.86 - 7.84$~ days \citep{Gompertz+2018, Rossi+2020, Rastinejad+2021, Jin+2015, Yang+2015}; GRB\,070714B: $\delta t \approx 1 - 1.03$~days \citep{Rossi+2020, Rastinejad+2021}; GRB\,070809: $\delta t \approx 0.47 - 1.46$~days \citep{Rossi+2020, Rastinejad+2021, Jin+2020}; GRB\,080503: $\delta t \approx 0.04 - 5.36$~days \citep{Rastinejad+2021, Zhou+2023}; GRB\,130603B: $\delta t \approx 9.37 - 9.49$~days \citep{Gompertz+2018, Rossi+2020, Rastinejad+2021, Berger+2013, Tanvir+2013}; GRB\,150101B: $\delta t \approx 1.66 - 10.71$~days \citep{Rossi+2020, Rastinejad+2021, Fong+2016, Troja+2018}; GRB\,160821B: $\delta t \approx 1.06 - 10.53$~days \citep{Rossi+2020, Rastinejad+2021, Kasliwal+2017, Lamb+2019}; GRB\,200522A: $\delta t \approx 3.12 - 3.66$~days \citep{Rossi+2020, O'Conner+2022}; GRB\,211211A: $\delta t \approx 0.3 - 5$~days \citep{Rastinejad+2022, Troja+2022, Yang+2022}; GRB\,230307A: $\delta t \approx 1 - 61$~days \citep{Gillanders+2023, Levan+2024, Yang+2024}.} For GRB\,211211A, we use an afterglow model from \citet{Rastinejad+2025} to estimate the afterglow contribution.}
\end{deluxetable*}

\section{X-ray and Optical $L_c$ with Environmental Properties}
\label{app:tables2}
We present a table showing our determined X-ray and optical luminosities at a common rest-frame time of $\delta t_c=3$~hr ($L_c$) for each GRB (see Section \ref{sec:methods}), along with its environmental properties from the Broadband Repository for Investigating Gamma-ray burst Host Traits (BRIGHT\footnote{\url{http://bright.ciera.northwestern.edu}}; \citealt{Fong+2022, Nugent+2022}) and from \citet{Levan+2024, Schroeder+2025, Nugent+2025}. Optical $L_c$s without $\sigma_{L_{C\, \text{opt}}}$ represent the deepest upper limit within 2.5 hr from $\delta t_c$.

\begin{longrotatetable}
\begin{deluxetable*}{ccccccccccccc}

\tabletypesize{\footnotesize}
\tablewidth{0pt}

\tablecaption{Afterglow Luminosities at a Common Rest-frame Time of 3~hr and Environmental Properties \label{tab:L_crf_results}}

\tablehead{
\colhead{GRB} &  
\colhead{z} &
\colhead{log($L_{C,{\rm opt}})$} & 
\colhead{log($\sigma_{L_{C,{\rm opt}}})$} & 
\colhead{log($L_{C,{\rm X}})$} & 
\colhead{log($\sigma_{L_{C,{\rm X}}})$} & 
\colhead{log($M_*/M_{\odot}$)} & 
\colhead{SFR} & 
\colhead{log(sSFR)} & 
\colhead{$\delta R_{{\rm phys}}$} & 
\colhead{$\delta R_{\rm norm}$} & 
\colhead{$A_V,{\rm host}$} & 
\colhead{$t_m$} \\
\colhead{} & 
\colhead{} &
\colhead{[erg/s]} & 
\colhead{[erg/s]} & 
\colhead{[erg/s]} & 
\colhead{[erg/s]} & 
\colhead{ } & 
\colhead{[$M_\odot$~yr$^{-1}$]} & 
\colhead{[yr$^{-1}$]} & 
\colhead{[kpc]} & 
\colhead{[$r_{e}$]} & 
\colhead{[mag]} & 
\colhead{[Gyr]}
}

\startdata
050509B & 0.225 & $<41.745$ & \nodata & 41.827 & 41.128 & 11.46 & 0.21 & -12.138 & 55.19 & 2.59 & \nodata & 8.84 \\
050709 & 0.161 & 42.91 & 42.09 & 44.464 & 43.765 & 8.55 & 0.02 & -10.249 & 3.76 & 2.0 & 0.02 & 0.57 \\
050724A & 0.254 & 43.74 & 42.91 & 44.094 & 43.395 & 11.05 & 0.15 & -11.874 & 2.74 & 0.67 & 0.76 & 8.20 \\
050813 & 0.719 & \nodata & \nodata & 42.209 & 41.51 & 10.31 & 1.53 & -10.125 & 43.57 & \nodata & 0.23 & 3.73 \\
051210 & 0.64 & \nodata & \nodata & 44.125 & 43.426 & 10.96 & 434.96 & -8.322 & 29.08 & 5.65 & 1.9 & 0.23 \\
051221A & 0.546 & 43.977 & 43.274 & 45.325 & 44.626 & 9.31 & 0.71 & -9.459 & 2.08 & 0.89 & 0.32 & 0.49 \\ \hline
060121 & 0.64 & 43.15 & 42.412 & 45.689 & 44.99 & \nodata & \nodata & \nodata & 0.97 & 0.18 & \nodata & \nodata \\
060313 & 0.64 & 43.969 & 43.453 & 45.296 & 44.597 & \nodata & \nodata & \nodata & 2.6 & 1.39 & \nodata & \nodata \\
060502B & 1.51 & 46.585 & 44.637 & \nodata & \nodata & \nodata & \nodata & \nodata & \nodata & \nodata & \nodata & \nodata \\
060614 & 0.125 & 41.65 & 41.89 & 44.63 & 43.94 & 7.85 & 0.004 & -10.25 & 0.7 & 0.86 & 0.1 & 0.76 \\
060801 & 1.13 & \nodata & \nodata & 44.377 & 43.678 & 9.12 & 9.18 & -8.157 & 10.25 & \nodata & 0.3 & 0.13 \\
061006 & 0.461 & 43.91 & 43.086 & 44.389 & 43.69 & 9.37 & 0.1 & -10.37 & 1.39 & 0.37 & 0.24 & 4.27 \\
061201 & 0.64 & 43.397 & 42.573 & 45.199 & 44.5 & \nodata & \nodata & \nodata & 33.09 & 14.91 & \nodata & \nodata \\
061210 & 0.41 & $<42.738$ & \nodata & 44.939 & 44.241 & 9.77 & 0.19 & -10.491 & 15.51 & \nodata & 0.3 & 0.66 \\
061217 & 0.827 & $<43.515$ & \nodata & \nodata & \nodata & \nodata & \nodata & \nodata & \nodata & \nodata & \nodata & \nodata \\ \hline
070429B & 0.902 & $<43.711$ & \nodata & 44.324 & 43.625 & 10.44 & 8.66 & -9.502 & 6.0 & 1.17 & 2.0 & 0.43 \\
070707 & 0.64 & 43.727 & 42.898 & \nodata & \nodata & \nodata & \nodata & \nodata & 3.25 & 1.11 & \nodata & \nodata \\
070714B & 0.923 & 44.601 & 43.384 & 44.884 & 44.185 & 9.37 & 1.23 & -9.28 & 12.33 & 5.17 & 0.33 & 1.63 \\
070724A & 0.457 & 42.603 & 41.779 & 44.478 & 43.779 & 9.81 & 6.48 & -8.998 & 5.52 & 1.49 & 1.25 & 0.27 \\
070729 & 0.52 & \nodata & \nodata & 43.744 & 43.045 & 8.76 & 0.89 & -8.811 & 19.72 & \nodata & 0.31 & 0.55 \\
070809 & 0.473 & 43.20 & 42.37 & 44.791 & 44.092 & 10.82 & 0.83 & -10.901 & 34.11 & 9.34 & 1.05 & 0.84 \\
071112B & 0.64 & $<43.241$ & \nodata & \nodata & \nodata & \nodata & \nodata & \nodata & \nodata & \nodata & \nodata & \nodata \\
071227 & 0.381 & 43.145 & 42.321 & 43.791 & 43.092 & 10.79 & 5.8 & -10.027 & 14.74 & 3.08 & 1.45 & 1.79 \\ \hline
080123 & 0.495 & \nodata & \nodata & 43.798 & 43.099 & 10.12 & 9.3 & -9.152 & 53.63 & \nodata & 0.62 & 0.43 \\
080426 & 1.7 & $<44.575$ & \nodata & 45.531 & 44.832 & \nodata & \nodata & \nodata & \nodata & \nodata & \nodata & \nodata \\
080503 & 0.64 & \nodata & \nodata & 43.887 & 43.188 & \nodata & \nodata & \nodata & 7.31 & 3.46 & \nodata & \nodata \\
080702A & 0.64 & \nodata & \nodata & 41.033 & 44.099 & \nodata & \nodata & \nodata & \nodata & \nodata & \nodata & \nodata \\
080905A & 0.1218 & 41.211 & 40.387 & 43.536 & 42.837 & \nodata & \nodata & \nodata & 18.3 & 4.61 & \nodata & \nodata \\
080919 & 0.64 & 46.05 & 45.226 & 43.778 & 43.079 & \nodata & \nodata & \nodata & \nodata & \nodata & \nodata & \nodata \\
081024A & 0.64 & $<44.829$ & \nodata & 43.872 & 43.173 & \nodata & \nodata & \nodata & \nodata & \nodata & \nodata & \nodata \\
081226A & 0.64 & 42.714 & 41.89 & 43.293 & 42.594 & \nodata & \nodata & \nodata & \nodata & \nodata & \nodata & \nodata \\ \hline
090305 & 0.64 & 42.892 & 41.446 & \nodata & \nodata & \nodata & \nodata & \nodata & \nodata & \nodata & \nodata & \nodata \\
090426 & 2.609 & 46.041 & 43.604 & 46.029 & 45.33 & \nodata & \nodata & \nodata & 0.49 & 0.29 & \nodata & \nodata \\
090510 & 0.903 & 43.975 & 43.481 & 44.914 & 44.215 & 9.75 & 1.26 & -9.65 & 10.51 & 1.66 & 0.54 & 0.45 \\
090515 & 0.403 & 41.511 & 40.688 & 43.388 & 42.689 & 11.25 & \nodata & \nodata & 76.19 & 13.98 & 0.1 & 6.34 \\
090607 & 0.64 & \nodata & \nodata & 43.402 & 42.703 & \nodata & \nodata & \nodata & \nodata & \nodata & \nodata & \nodata \\
090621B & 0.64 & \nodata & \nodata & 44.588 & 43.889 & \nodata & \nodata & \nodata & \nodata & \nodata & \nodata & \nodata \\
090916 & 0.64 & $<44.794$ & \nodata & \nodata & \nodata & \nodata & \nodata & \nodata & \nodata & \nodata & \nodata & \nodata \\
091109B & 0.64 & 42.81 & 41.986 & 44.542 & 43.843 & \nodata & \nodata & \nodata & 4.22 & 1.93 & \nodata & \nodata \\ \hline
100117A & 0.914 & 42.918 & 42.094 & 44.043 & 43.344 & 10.35 & \nodata & \nodata & 1.35 & 0.61 & 0.16 & 3.02 \\
100206A & 0.407 & \nodata & \nodata & 42.697 & 41.998 & 10.72 & 7.64 & -9.837 & 25.28 & \nodata & 1.19 & 4.58 \\
100213A & 0.64 & \nodata & \nodata & 44.208 & 43.509 & \nodata & \nodata & \nodata & \nodata & \nodata & \nodata & \nodata \\
100625A & 0.452 & \nodata & \nodata & 42.816 & 42.117 & 9.7 & \nodata & \nodata & 2.63 & \nodata & 0.1 & 3.56 \\
100628A & 0.64 & $<42.785$ & \nodata & \nodata & \nodata & \nodata & \nodata & \nodata & \nodata & \nodata & \nodata & \nodata \\
100702A & 0.64 & $<43.232$ & \nodata & 44.609 & 43.91 & \nodata & \nodata & \nodata & \nodata & \nodata & \nodata & \nodata \\
101219A & 0.718 & $<42.986$ & \nodata & 42.215 & 44.00 & 9.39 & 3.14 & -8.893 & 5.48 & \nodata & 1.5 & 0.25 \\
101224A & 0.454 & \nodata & \nodata & 43.904 & 43.205 & 9.17 & 0.58 & -9.407 & 12.75 & \nodata & 0.2 & 0.46 \\ \hline
110112A & 0.64 & 43.798 & 42.974 & 44.592 & 43.893 & \nodata & \nodata & \nodata & 18.03 & 6.53 & \nodata & \nodata \\
111020A & 0.64 & \nodata & \nodata & 45.154 & 44.455 & \nodata & \nodata & \nodata & \nodata & \nodata & \nodata & \nodata \\
111117A & 2.211 & $<44.268$ & \nodata & 45.019 & 44.32 & 9.63 & 22.11 & -8.285 & 10.52 & \nodata & 0.3 & 0.19 \\
111121A & 0.64 & \nodata & \nodata & 45.452 & 44.753 & \nodata & \nodata & \nodata & \nodata & \nodata & \nodata & \nodata \\ \hline
120305A & 0.225 & \nodata & \nodata & 43.1 & 42.401 & 9.17 & 0.03 & -10.693 & 18.09 & \nodata & 0.02 & 2.11 \\
120521A & 0.64 & \nodata & \nodata & 44.686 & 43.987 & \nodata & \nodata & \nodata & \nodata & \nodata & \nodata & \nodata \\
120804A & 1.05 & 43.319 & 42.495 & 45.781 & 45.082 & 9.81 & 18.79 & -8.536 & 2.22 & \nodata & 2.77 & 0.35 \\
121226A & 1.37 & $<44.436$ & \nodata & 45.706 & 45.007 & 9.46 & 24.59 & -8.069 & 2.31 & \nodata & 1.01 & 0.12 \\ \hline
130313A & 2.79 & $<44.161$ & \nodata & \nodata & \nodata & \nodata & \nodata & \nodata & \nodata & \nodata & \nodata & \nodata \\
130515A & 0.8 & \nodata & \nodata & 43.979 & 43.28 & 10.28 & 0.26 & -10.865 & 61.22 & \nodata & 0.14 & 0.78 \\
130603B & 0.357 & 43.803 & 43.044 & 45.236 & 44.537 & 9.82 & 0.44 & -10.177 & 5.4 & 0.71 & 0.29 & 1.63 \\
130716A & 2.2 & \nodata & \nodata & 44.832 & 44.133 & 9.61 & 11.99 & -8.531 & 33.08 & \nodata & 0.4 & 0.32 \\
130822A & 0.154 & \nodata & \nodata & 41.754 & 41.055 & 10.16 & 0.3 & -10.683 & 60.09 & \nodata & 0.21 & 2.16 \\
130912A & 0.64 & 43.66 & 42.835 & 45.127 & 44.428 & \nodata & \nodata & \nodata & 3.9 & 1.41 & \nodata & \nodata \\
131004A & 0.717 & 44.134 & 43.477 & 45.005 & 44.306 & \nodata & \nodata & \nodata & 0.8 & 0.25 & \nodata & \nodata \\ \hline
140129B & 0.43 & 45.513 & 43.384 & 44.769 & 44.07 & 9.33 & 0.06 & -10.552 & 1.76 & \nodata & 0.09 & 1.65 \\
140320A & 0.64 & \nodata & \nodata & 47.081 & 46.382 & \nodata & \nodata & \nodata & \nodata & \nodata & \nodata & \nodata \\
140516A & 1.47 & $<43.533$ & \nodata & 43.606 & 42.907 & \nodata & \nodata & \nodata & \nodata & \nodata & \nodata & \nodata \\
140606A & 0.64 & $<43.75$ & \nodata & \nodata & \nodata & \nodata & \nodata & \nodata & \nodata & \nodata & \nodata & \nodata \\
140622A & 0.959 & \nodata & \nodata & 43.404 & 42.705 & 10.17 & 6.01 & -9.391 & 32.95 & \nodata & 0.61 & 0.66 \\
140903A & 0.353 & 44.733 & 43.909 & 45.409 & 44.71 & 10.81 & 2.28 & -10.452 & 0.9 & \nodata & 2.99 & 4.24 \\
140930B & 1.465 & 44.104 & 43.28 & 45.325 & 44.626 & 9.45 & 5.13 & -8.74 & 9.62 & \nodata & 0.28 & 0.57 \\
141212A & 0.596 & $<43.942$ & \nodata & 43.979 & 43.28 & 9.71 & 1.17 & -9.642 & 18.75 & \nodata & 0.43 & 2.37 \\ \hline
150101A & 0.64 & \nodata & \nodata & 43.285 & 42.586 & \nodata & \nodata & \nodata & \nodata & \nodata & \nodata & \nodata \\
150101B & 0.134 & \nodata & \nodata & 44.409 & 43.71 & 11.13 & 0.22 & -11.788 & 7.36 & 0.78 & 0.25 & 4.88 \\
150120A & 0.4604 & $<42.021$ & \nodata & 43.915 & 43.216 & 10.01 & 2.27 & -9.654 & 4.77 & \nodata & 1.1 & 2.28 \\
150423A & 1.394 & 43.664 & 42.365 & 43.429 & 42.73 & \nodata & \nodata & \nodata & \nodata & \nodata & \nodata & \nodata \\
150424A & 0.64 & 43.449 & 42.625 & 45.812 & 45.113 & \nodata & \nodata & \nodata & 3.41 & 1.5 & \nodata & \nodata \\
150728A & 0.461 & \nodata & \nodata & 44.754 & 44.055 & 9.35 & 8.13 & -8.44 & 7.52 & \nodata & 0.86 & 9.35 \\
150831A & 1.18 & \nodata & \nodata & 45.014 & 44.315 & 9.49 & 5.97 & -8.714 & 12.43 & \nodata & 1.02 & 0.51 \\
151229A & 0.63 & $<45.23$ & \nodata & 46.168 & 45.469 & 8.79 & 0.22 & -9.448 & 8.16 & \nodata & 0.87 & 1.78 \\ \hline
160303A & 1.01 & 43.891 & 43.127 & 45.152 & 44.453 & 9.51 & 2.37 & -9.135 & 15.31 & 3.42 & 1.61 & 1.05 \\
160408A & 1.9 & $<44.522$ & \nodata & 44.949 & 44.25 & 9.32 & 3.57 & -8.767 & 14.13 & \nodata & 0.49 & 0.62 \\
160410A & 1.7177 & 44.897 & 42.843 & 45.865 & 45.166 & \nodata & \nodata & \nodata & \nodata & \nodata & \nodata & \nodata \\
160411A & 0.82 & $<43.444$ & \nodata & 44.565 & 43.866 & 8.87 & 1.09 & -8.833 & 1.4 & \nodata & 0.68 & 0.67 \\
160525B & 0.64 & $<43.509$ & \nodata & 43.218 & 42.519 & 8.04 & 1.38 & -7.9 & 5.5 & \nodata & 0.14 \\ 
160601A & 0.64 & 43.423 & 42.475 & 44.472 & 43.773 & \nodata & \nodata & \nodata & 1.38 & \nodata & \nodata & \nodata \\ 
160624A & 0.484 & $<44.396$ & \nodata & 44.837 & 44.138 & 9.74 & 1.28 & -9.633 & 9.63 & 2.37 & 0.55 & 1.19 \\ 
160821B & 0.162 & 41.677 & 40.853 & 43.841 & 43.142 & 9.24 & 0.24 & -9.86 & 15.74 & 4.24 & 0.01 & 0.58 \\ 
160927A & 0.64 & 42.248 & 41.424 & 44.549 & 43.85 & \nodata & \nodata & \nodata & \nodata & \nodata & \nodata & \nodata \\ 
161001A & 0.67 & \nodata & \nodata & 45.831 & 45.132 & 9.73 & 0.53 & -10.006 & 18.54 & \nodata & 0.45 & 0.78 \\ 
161104A & 0.793 & \nodata & \nodata & 43.792 & 43.093 & 10.23 & 0.06 & -11.452 & 1.66 & \nodata & 0.08 & 2.26 \\ \hline
170127B & 2.21 & $<44.675$ & \nodata & 45.107 & 44.408 & 9.51 & 10.25 & -8.499 & 10.37 & \nodata & 0.58 & 0.31 \\ 
170428A & 0.453 & 42.43 & 41.606 & 44.561 & 43.862 & 9.68 & 0.4 & -10.078 & 7.72 & \nodata & \nodata & 5.14 \\
170728A & 1.493 & \nodata & \nodata & 45.735 & 45.037 & 10.09 & 79.83 & -8.188 & 32.25 & \nodata & 1.72 & 0.16 \\
170728B & 1.272 & 44.242 & 43.418 & 46.995 & 46.296 & 9.87 & 10.88 & -8.833 & 8.4 & \nodata & 0.42 & 0.41 \\
170817 & 0.01 & \nodata & \nodata & \nodata & \nodata & 10.61 & 0.02 & -12.309 & 2.125 & 0.57 & 0.03 & 10.42 \\ \hline
180402A & 0.64 & $<43.18$ & \nodata & 44.428 & 43.729 & \nodata & \nodata & \nodata & \nodata & \nodata & \nodata & \nodata \\ 
180418A & 1.56 & 45.108 & 43.146 & 45.982 & 45.283 & 9.83 & 13.0 & -8.716 & 1.3 & \nodata & 1.3 & 0.56 \\ 
180618A & 0.52 & 43.772 & 42.948 & 44.683 & 43.984 & 8.81 & 1.87 & -8.538 & 9.7 & \nodata & 0.32 & 0.35 \\ 
180727A & 1.95 & $<46.184$ & \nodata & 45.391 & 44.692 & 9.23 & 3.07 & -8.743 & 2.56 & \nodata & 0.68 & 0.54 \\ 
180805B & 0.661 & \nodata & \nodata & 44.489 & 43.79 & 9.34 & 1.53 & -9.155 & 24.3 & \nodata & 0.28 & 0.50 \\ 
181123B & 1.754 & 43.581 & 42.757 & 44.951 & 44.252 & 9.9 & 11.72 & -8.831 & 5.08 & \nodata & 0.36 & 0.63 \\ \hline
191031D & 1.93 & $<44.783$ & \nodata & 44.751 & 44.053 & 10.38 & 36.14 & -8.822 & 13.08 & \nodata & 1.41 & 0.8 \\ \hline
200219A & 0.48 & $<44.757$ & \nodata & 43.012 & 42.313 & 10.74 & 9.91 & -9.744 & 8.3 & \nodata & 0.96 & 3.52 \\ 
200411A & 0.82 & \nodata & \nodata & 45.364 & 44.665 & 10.23 & 27.35 & -8.793 & 37.66 & \nodata & 1.52 & 0.62 \\ 
200522A & 0.554 & $<43.471$ & \nodata & 44.694 & 43.995 & 9.66 & 2.23 & -9.312 & 0.93 & 0.24 & 0.01 & 0.58 \\ 
200907B & 0.56 & \nodata & \nodata & 44.006 & 43.307 & 9.35 & 1.32 & -9.229 & 2.41 & \nodata & 0.59 & 0.89 \\ 
201006A & 0.64 & \nodata & \nodata & 43.793 & 43.094 & \nodata & \nodata & \nodata & \nodata & \nodata & \nodata & \nodata \\ 
201221D & 1.055 & 43.455 & 42.631 & 43.299 & 42.6 & 9.36 & 2.36 & -8.987 & 29.35 & \nodata & 0.17 & 0.27 \\ \hline
210323A & 0.733 & 42.916 & 41.714 & 45.466 & 44.767 & 8.77 & 0.34 & -9.239 & 5.89 & \nodata & 0.07 & 0.56 \\ 
210726A & 0.37 & $<42.968$ & \nodata & 44.31 & 43.611 & 7.84 & 0.06 & -9.062 & 0.23 & \nodata & 0.68 & 1.06 \\ 
210919A & 0.242 & $<42.424$ & \nodata & 41.846 & 41.147 & 9.87 & 0.3 & -10.393 & 51.05 & \nodata & 0.8 & 1.62 \\ 
211023B & 0.862 & 45.137 & 44.313 & 45.222 & 44.523 & 9.65 & 1.44 & -9.492 & 3.84 & \nodata & 0.42 & 1.72 \\ 
211106A & 0.64 & $<42.5$ & \nodata & 45.461 & 44.762 & \nodata & \nodata & \nodata & 0.79 & 0.49 & \nodata & \nodata \\ 
211211A & 0.0763 & 42.32 & 41.49 & 44.453 & 43.754 & 8.84 & 0.07 & -9.995 & 7.92 & 3.2 & 0.05 & 2.53 \\ \hline
220412B & 0.64 & $<44.447$ & \nodata & \nodata & \nodata & \nodata & \nodata & \nodata & \nodata & \nodata & \nodata & \nodata \\ 
221120A & 0.64 & $<43.855$ & \nodata & \nodata & \nodata & \nodata & \nodata & \nodata & \nodata & \nodata & \nodata & \nodata \\ \hline
230205A & 0.429 & \nodata & \nodata & 45.069 & 44.37 & \nodata & \nodata & \nodata & \nodata & \nodata & \nodata & \nodata \\ 
230217A & 0.64 & \nodata & \nodata & 45.715 & 45.016 & \nodata & \nodata & \nodata & \nodata & \nodata & \nodata & \nodata \\ 
230228A & 0.64 & $<44.06$ & \nodata & 44.822 & 44.123 & \nodata & \nodata & \nodata & \nodata & \nodata & \nodata & \nodata \\ 
230307A & 0.065 & 44.51 & 43.69 & 43.81 & 43.11 & 9.66 & 0.09 & -10.71 & 38.9 & 12.1 & 0.27 & 5.5 \\ 
231117A & 0.257 & 43.543 & 41.929 & 44.45 & 43.75 & 9.16 & 0.39 & -9.57 & 2.0 & \nodata & 0.82 & 1.62 \\ 
\enddata

\tablecomments{Environmental properties were taken from the Broadband Repository for Investigating Gamma-ray burst Host Traits (BRIGHT; \citealt{Fong+2022, Nugent+2022}) and from \citet{Levan+2024, Schroeder+2025, Nugent+2025}}

\end{deluxetable*}
\end{longrotatetable}

\end{document}